\documentclass[11pt]{article}
\pdfoutput=1
\usepackage{jheppub}
\usepackage{tabu}
\usepackage[vcentermath]{youngtab}
\usepackage[usenames,dvipsnames,table]{xcolor}
\usepackage{graphicx,amsmath,amssymb,amsthm,multirow,array,bm,bbm,esint}
\usepackage[mathscr]{eucal}
\usepackage[bbgreekl]{mathbbol}
\usepackage{epsf,amsfonts}
\usepackage{slashed}
\usepackage[numbers,sort&compress]{natbib}
\usepackage{minibox}
\usepackage{rotating}
\usepackage{pdflscape}
\usepackage{array,tikz-cd}
\usepackage{subcaption}
\usepackage{framed}
\usepackage{mathtools}
\usepackage{tikz} 
\usepackage[compat=1.1.0]{tikz-feynman}
\usepackage{feynmf}
\usepackage{float}
\usepackage{verbatim}

\usepackage{slashed}

\def \tr{\mbox{tr\,}}

\newcommand{\ads}{\text{\tiny AdS} }

\newcommand{\im}{\text{Im}}

\newcommand{\const}{\text{\tiny BTZ} }

\newcommand{\kc}{k_{\tiny cs} }	

\newcommand{\ons}{\text{\tiny on-shell} }	

\newcommand{\be}{\begin{equation}}
\newcommand{\ee}{\end{equation}}
\newcommand{\bea}{\begin{eqnarray}}
\newcommand{\eea}{\end{eqnarray}}

\newcommand{\mt}[1]{\textrm{\tiny #1}}

\newcommand{\vev}[1]{\langle #1\rangle}

\newcommand{\ie}{\emph{i.e.}, }
\newcommand{\eg}{\emph{e.g.}, }

\makeatletter
\newcommand{\doubletilde}[1]{{%
  \mathpalette\double@tilde{#1}%
}}
\newcommand{\double@tilde}[2]{%
  \sbox\z@{$\m@th#1\tilde{#2}$}%
  \ht\z@=.9\ht\z@
  \tilde{\box\z@}%
}
\makeatother

\title{On the Chaos Bound in Rotating Black Holes}

\author[a]{Viktor Jahnke}
\author[b]{\!, Keun-Young Kim}
\author[c]{\!, Junggi Yoon}

\affiliation[\,a,b]{School of Physics and Chemistry, Gwangju Institute of Science and Technology, 123 Cheomdan-gwagiro, Gwangju 61005, Korea}
\affiliation[\,c]{School of Physics, Korea Institute for Advanced Study\\
85 Hoegiro Dongdaemun-gu, Seoul 02455, Republic of Korea}

\emailAdd{viktorjahnke@gist.ac.kr}
\emailAdd{fortoe@gist.ac.kr}
\emailAdd{junggiyoon@kias.re.kr}

\preprint{{\raggedleft \tt KIAS-P19018 \par} }

\abstract{ We study out-of-time-order correlators (OTOCs) of rotating BTZ black holes using two different approaches: the elastic eikonal gravity approximation, and the Chern-Simons formulations of 3-dimensional gravity. Within both methods the OTOC is given as a sum of two contributions, corresponding to left and right moving modes. The contributions have different Lyapunov exponents, $\lambda_L^{\pm}=\frac{2\pi}{\beta}\frac{1}{1\mp \ell \Omega}$, where $\Omega$ is the angular velocity and $\ell$ is the AdS radius. Since $\lambda_L^{-} \leq \frac{2\pi}{\beta} \leq \lambda_L^{+}$, there is an apparent contradiction with the chaos bound. We discuss how the result can be made consistent with the chaos bound if one views the parameters $\beta_{\pm}=\beta(1\mp \ell \Omega)$ as the effective inverse temperatures of the left and right moving modes. 
}

\begin{document}
\maketitle

\section{Introduction}
\label{sec:introduction}

In recent years, the gauge-gravity duality~\cite{Maldacena:1997re,Gubser:1998bc,Witten:1998qj} has provided remarkable insights into the nature of quantum chaos\footnote{These recent developments were reviewed in~\cite{Sarosi:2017ykf,Jahnke:2018off}.}. An important lesson from the holographic approach to quantum chaos~\cite{Shenker:2013pqa, Shenker:2013yza,Roberts:2014isa,Shenker:2014cwa} is the fact that the onset of chaos can be efficiently diagnosed with the so-called out-of-time-order correlators~(OTOCs)
\be
F(t,\vec{x})= \langle V(0) W(t,\vec{x}) V(0) W(t,\vec{x}) \rangle\,,
\ee
which, for chaotic systems, are expected to vanish at later times for almost any local operators $V$ and $W$. In holographic systems at finite temperature, the OTOC has a very simple form
\be
\frac{\langle V(0) W(t,\vec{x}) V(0) W(t,\vec{x}) \rangle}{\langle V(0) V(0)\rangle \langle W(t,\vec{x}) W(t,\vec{x}) \rangle} = 1-\varepsilon_{\Delta_V \Delta_W} \exp \left[\lambda_L\left(t-t_*-\frac{|\vec{x}|}{v_B} \right) \right]\,,
\ee
where the multiplicative factor, $\varepsilon_{\Delta_V \Delta_W}$, contains information about the operators $V$ and $W$, while the exponential corresponds to an universal piece, and it is characterized by only three parameters: the Lyapunov exponent $\lambda_L$, the scrambling time $t_*$, and the butterfly velocity $v_B$. All these quantities are determined from the geometry close to the black hole horizon. For black holes, one generically expects $\lambda_L = \frac{2\pi}{\beta}$, where $\beta$ is the system's inverse temperature. Surprisingly, the above value provides an upper bound (the {\it chaos bound}) for the Lyapunov exponent in generic large-$N$ systems, namely \cite{Maldacena:2015waa}
\be
\lambda_L \leq \frac{2\pi}{\beta}\,.
\label{eq-chaosBound}
\ee
This highlights a special property of holographic systems which are dual to black holes - they saturate the chaos bound. This finding generated a lot of excitement in the community, leading to the speculation that the saturation of the chaos bound could be a sufficient condition for the existence of Einstein gravity dual~\cite{Maldacena:2015waa}. However, despite being a necessary condition, the saturation of the chaos does not guarantee the existence of dual description purely in terms of Einstein gravity~\cite{deBoer:2017xdk,Banerjee:2018twd,Banerjee:2018kwy}. The observation that the SYK model saturates the chaos bound put this system into evidence as a prototypical example of a simple model for holography. Several aspects of this system can be capture by a 2-dimensional Einstein-dilation theory, and this has generated a lot of research within the framework of NAdS$_2$/NCFT$_1$ and related areas. See, for instance, \cite{Sachdev:1992fk,Polchinski:2016xgd,Jevicki:2016bwu,Maldacena:2016hyu,Maldacena:2016upp,Jevicki:2016ito,Gross:2016kjj,Fu:2016vas,Ahn:2018sgn,Garcia-Garcia:2019poj,deMelloKoch:2019ywq,Kim:2019upg,Ferrari:2019ogc}.

One expects the saturation of the chaos bound to be a property of quite generic black holes~\cite{Maldacena:2015waa}. However, some recent studies suggest that rotating black holes have two Lyapunov exponents, one of which does not saturate the chaos bound, while the other violates it~\cite{Poojary:2018esz,Stikonas:2018ane}. Both these works consider the 3-dimensional rotating BTZ black hole. In \cite{Poojary:2018esz}, the author finds an effective action for the boundary degrees of freedom and studies OTOCs for this effective theory. He finds that the OTOC is controlled by two modes (left and right moving modes), with Lyapunov exponents $\lambda_L^{\pm}=\frac{2\pi}{\beta}\frac{1}{1\mp \ell \Omega}$, where $\ell \Omega$ is the chemical potential for angular momentum. In \cite{Stikonas:2018ane}, the author studies the disruption of the two-sided mutual information both in the CFT and in the bulk. In his particular configuration, the onset of chaos is controlled by $\lambda_L^{-}$, which indicates a non-saturation of the chaos bound. Moreover, the aforementioned works seem to be in contradiction with \cite{Reynolds:2016pmi}, in which the author studies chaos for the rotating BTZ black holes using the geodesic approximation and finds a saturation of the chaos bound.

In this paper, we revisit the calculation of OTOCs for rotating BTZ black holes using two different methods. We first use the elastic eikonal gravity approximation \cite{Shenker:2014cwa}, in which the OTOCs have a very vivid holographic representation in terms of a high energy shock wave collision near the black hole's bifurcation surface. We then study chaos using the Chern-Simons formulation of 3-dimensional gravity. We show that the dynamics of the boundary degrees of freedom is governed by two copies of a Schwarzian-like action and derive the OTOCs from the analytic continuation of the euclidean 4-point. We compare both approaches with each other and with previous results in the literature, and discuss an apparent violation of the chaos bound in this system.

The paper is organized as follows. In Section~\ref{sec: review} we review the rotating BTZ black hole geometry and discuss a few aspects of the corresponding CFT dual description. In Section~\ref{sec: eikonal approximation} we compute OTOCs using the gravity eikonal approximation. In Section \ref{sec: cs gravity review} we review the Chern-Simons formulation of 3-dimensional gravity and we derive an effective action for the boundary degrees of freedom. We then evaluate OTOCs from the on-shell action via analytic continuation. We discuss our results in Section~\ref{sec: discussion} and relegate some technical details to the Appendix \ref{app: finite transf}.

\section{The rotating BTZ black hole}
\label{sec: review}

In this section we briefly review the rotating BTZ geometry~\cite{Banados:1992wn,Banados:1992gq}. This is a solution of Einstein gravity in 2+1 dimensions with a negative cosmological constant $\Lambda=-1/\ell^2$. In terms of {\it Schwarzschild coordinates} $(t,r,\varphi)$, the metric reads
\be
ds^2=- f(r) dt^2+\frac{dr^2}{f(r)}+r^2 \left( d\varphi -\frac{r_{+}r_{-}}{\ell r^2}dt\right)^2\,,
\label{eq-metricS}
\ee
where $\varphi$ is periodic with period $2\pi$ and the blackening factor has the form
\be
f(r)=\frac{(r^2-r_{+}^2)(r^2-r_{-}^2)}{\ell^2 r^2}\,.
\ee
Here $r_{+}$ and $r_{-}$ are the radii of the outer and inner horizons, respectively. The black hole's mass $M$, angular momentum $J$, and temperature $T$ are determined from $r_{\pm}$
\be
M=\frac{r_{+}^2+r_{-}^2}{\ell^2}\,,\,\,\,\,\,\,\,\,J=\frac{2\, r_{+} r_{-}}{\ell}\,,\,\,\,\,\,\,\,\, \beta=\frac{1}{T}=\frac{2\pi \ell^2 r_{+}}{r_{+}^2-r_{-}^2}\,.
\ee

Since we will be interested in the near-horizon geometry, it is convenient to work with {\it co-rotating coordinates} $(t,r,\phi)$, with the new angular variable defined as
\be
\phi\equiv \varphi-\Omega t\,,
\ee
where 
\be
\Omega =\frac{r_{-}}{\ell r_{+}} \,,
\ee
is the angular velocity of the outer horizon. In these coordinates the metric takes the form
\be
ds^2=- f(r) dt^2+\frac{dr^2}{f(r)}+r^2 \left( N^{\phi}(r)dt+d\phi \right)^2\,,
\label{eq-metricCoRot}
\ee
where
\be
N^{\phi}(r)=\frac{r_{-}}{r_{+}}\frac{r^2-r_{+}^2}{\ell r^2}\,.
\ee
In the near-horizon limit, $N^{\phi}(r)$ vanishes and the metric becomes diagonal.

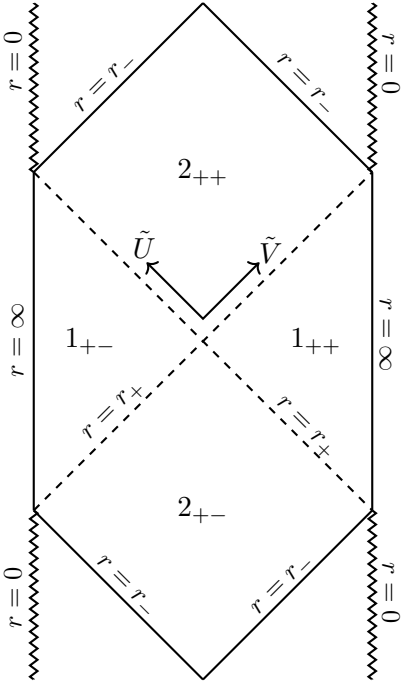
\begin{figure}[]
\centering

\begin{tikzpicture}[scale=1.5]
\draw [thick]  (0,0) -- (0,3);
\draw [thick]  (3,0) -- (3,3);
\draw [thick]  (0,3) -- (1.5,4.5);
\draw [thick]  (1.5,4.5) -- (3,3); 
\draw [thick,dashed]  (0,0) -- (3,3);
\draw [thick,dashed]  (0,3) -- (3,0);
\draw [thick] (0,0) -- (1.5,-1.5);
\draw [thick] (1.5,-1.5) -- (3,0);
\draw [thick,decorate,decoration={zigzag,segment length=1.5mm,amplitude=.5mm}] (0,3) -- (0,4.5);
\draw [thick,decorate,decoration={zigzag,segment length=1.5mm,amplitude=.5mm}] (0,0) -- (0,-1.5);
\draw [thick,decorate,decoration={zigzag,segment length=1.5mm,amplitude=.5mm}] (3,3) -- (3,4.5);
\draw [thick,decorate,decoration={zigzag,segment length=1.5mm,amplitude=.5mm}] (3,0) -- (3,-1.5);

\node[scale=1,align=center] at (2.5,1.5) {$1_{++}$};
\node[scale=1,align=center] at (0.5,1.5) {$1_{+-}$};
\node[scale=1, align=center] at (1.5,3) {$2_{++}$};
\node[scale=1, align=center] at (1.5,0) {$2_{+-}$};

\node[scale=1,align=center] at (0.98,2.36){$\tilde{U}$};
\node[scale=1,align=center] at (2.1,2.31){$\tilde{V}$};

\draw[thick,<->] (1,2.2) -- (1.5,1.7) -- (2,2.2);

\end{tikzpicture}
\put(-138,115){\rotatebox{90}{\small $r = \infty$}}
\put(-141,223){\rotatebox{90}{\small $r = 0$}}
\put(-141,20){\rotatebox{90}{\small $r = 0$}}
\put(1,145){\rotatebox{-90}{\small $r = \infty$}}
\put(2,245){\rotatebox{-90}{\small $r = 0$}}
\put(2,45){\rotatebox{-90}{\small $r = 0$}}
\put(-112,91){\rotatebox{45}{\small $r = r_{+}$}}
\put(-40,105){\rotatebox{-45}{\small $r = r_{+}$}}
\put(-48,25){\rotatebox{45}{\small $r = r_{-}$}}
\put(-108,45){\rotatebox{-45}{\small $r = r_{-}$}}
\put(-115,215){\rotatebox{45}{\small $r = r_{-}$}}
\put(-40,235){\rotatebox{-45}{\small $r = r_{-}$}}

\caption{ \small Penrose diagram for the rotating BTZ black hole. The figure only shows the region of interest. The same pattern repeats itself indefinitely above and below \cite{Banados:1992gq}. The regions $1_{++}$ and $1_{+-}$ denote the right and left exterior regions, while the regions $2_{+-}$ and $2_{++}$ denote the past and future interiors. The coordinates  $(\tilde{U},\tilde{V})$ are defined as in \cite{Fidkowski:2003nf}: $V=e^{\frac{\pi}{2}}\text{tan}\left( \frac{\tilde{V}}{2}\right)$ and $U=e^{\frac{\pi}{2}}\text{tan}\left( \frac{\tilde{U}}{2}\right)$, with $\tilde{U},\tilde{V} \in [-\pi,\pi]$.}
\label{fig-Penrose}
\end{figure}

Figure \ref{fig-Penrose} shows the Penrose diagram for the maximally extented BTZ geometry. Here we follow \cite{Balasubramanian:2004zu} and denote the left and right exterior regions as $1_{+-}$ and $1_{++}$, while the future and past interiors are denoted as $2_{++}$ and $2_{+-}$, respectively. In Figure \ref{fig-Penrose} we only shows the region of interest. The same structure repeats itself in the vertical direction.

For later purposes, we introduce Kruskal coordinates for the region $1_{++}$
\be
U=-e^{-\kappa(t-r_*)}\,,\,\,\,\,\,V=e^{\kappa(t+r_*)}\,,\,\,\,\,\, \kappa=\frac{r_{+}^2-r_{-}^2}{\ell^2 r_{+}}\,,
\ee
where the tortoise coordinate $r_*$ is \cite{Reynolds:2016pmi}
\be
r_*=\frac{1}{2 \kappa} \log \left( \frac{\sqrt{r^2-r_{-}^2}-\sqrt{r_{+}^2-r_{-}^2}}{\sqrt{r^2-r_{-}^2}+\sqrt{r_{+}^2-r_{-}^2}}\right)\,.
\ee
In these coordinates the metric takes the form 
\be
ds^2=\frac{-4\ell^2 dU dV-4\ell r_{-}(UdV-VdU)d\phi+\left[ (1-UV)^2r_{+}^2+4UV r_{-}^2\right]d\phi^2}{(1+UV)^2}
\label{eq-metricKruskal}
\ee

\subsection{Embedding coordinates and bulk-to-boundary propagators}
 The AdS$_3$ space is defined as the hyperboloid $-T_1^2-T_2^2+X_1^2+X_2^2=-\ell^2$ embedded in a space with metric $ds^2=-dT_1^2-dT_2^2+dX_1^2+dX_2^2$.  The rotating BTZ geometry can be thought of as piece of pure AdS$_3$, with identifications in the angular coordinates, namely $\varphi \sim \varphi + 2\pi$. This fact allows us to compute bulk-to-boundary propagators for this geometry. 

We start by defining embedding coordinates for the region 1$_{++}$. In terms of co-rotating coordinates, the embedding coordinates take the form
\bea
T_1 &=&\ell \,\sqrt{\frac{r^2-r_{+}^2}{r_{+}^2-r_{-}^2}} \sinh \left( \kappa t-\frac{r_{-}}{\ell}\phi \right)\,,\\
T_2 &=&\ell \, \sqrt{\frac{r^2-r_{-}^2}{r_{+}^2-r_{-}^2}} \cosh \left( \frac{r_{+}}{\ell}\phi \right)\,,\\
X_1 &=&\ell \, \sqrt{\frac{r^2-r_{+}^2}{r_{+}^2-r_{-}^2}} \cosh \left( \kappa t-\frac{r_{-}}{\ell}\phi \right)\,,\\
X_2 &=&\ell \, \sqrt{\frac{r^2-r_{-}^2}{r_{+}^2-r_{-}^2}} \sinh \left( \frac{r_{+}}{\ell}\phi \right)\,.
\eea
In terms of Kruskal coordinates, we have
\bea
T_1 &=&\ell \,\frac{V+U}{1+UV} \cosh \left( \frac{r_{-}}{\ell} \phi\right)-\ell \, \frac{V-U}{1+UV} \sinh \left( \frac{r_{-}}{\ell} \phi\right)\,,\\
T_2 &=&\ell \,\frac{1-UV}{1+UV} \cosh \left( \frac{r_{+}}{\ell} \phi\right) \,,\\
X_1 &=&\ell \,\frac{V-U}{1+UV} \cosh \left( \frac{r_{-}}{\ell} \phi\right)-\ell \, \frac{V+U}{1+UV} \sinh \left( \frac{r_{-}}{\ell} \phi\right)\,,\\
X_2 &=&\ell \,\frac{1-UV}{1+UV} \sinh \left( \frac{r_{+}}{\ell} \phi\right) \,.
\eea
By replacing the above formulas in the metric $ds^2=-dT_1^2-dT_2^2+dX_1^2+dX_2^2$, we can recover the formulas (\ref{eq-metricCoRot}) and (\ref{eq-metricKruskal}). 

Having defined the embedding coordinates, we can now compute the geodesic distance $d$ between two points $p=(T_1,T_2,X_1,X_2)$ and $p'=(T_1',T_2',X_1',X_2')$ as
\be
\cosh \left( \frac{d}{\ell}\right)=\frac{1}{\ell^2}\left( T_1 T_1' + T_2 T_2' - X_1 X_1' -X_2 X_2'\right) \,.
\ee
We will use the above formulas to compute bulk-to-boundary propagators. For our calculation, it will be convenient to write the boundary point in terms of co-rotating coordinates $p'=(t,r,\phi')$ and the bulk point in terms of Kruskal coordinates $p=(U,V,\phi)$. The geodesic distance between such points can be written as
\be
\cosh \left( \frac{d}{\ell}\right)=\frac{r_{\infty}(r_{+}^2-r_{-}^2)^{-1/2}}{1+UV} \left[U e^{\kappa t'+\frac{r_{-}}{\ell}\Delta \phi} -V e^{-\left(\kappa t'+\frac{r_{-}}{\ell}\Delta \phi \right)}+(1-UV) \cosh \left( \frac{r_{+}}{\ell} \Delta \phi \right)\right] \,,
\ee
where $\Delta \phi =\phi-\phi'$ and $r_{\infty}$ is an UV cutoff defining the radial position of the boundary point. 

From the above expression we can obtain the bulk-to-boundary propagator as proportional to $r_{\infty}^{\Delta}\left( \cosh \left( \frac{d}{\ell}\right) \right)^{-\Delta}$, i.e.,
\be
\langle \Phi (U,V,\phi) \mathcal{O}(t',\phi')\rangle=\frac{c}{1+UV} \left[U e^{\kappa t'+\frac{r_{-}}{\ell}\Delta \phi} -V e^{-\left(\kappa t'+\frac{r_{-}}{\ell}\Delta \phi \right)}+(1-UV) \cosh \left( \frac{r_{+}}{\ell} \Delta \phi \right)\right]^{-\Delta} \,,
\ee
where $c$ is a constant that depends on the operator $\mathcal{O}$. Since we are only going to compute the bulk-to-boundary propagators when the bulk point is either at $U=0$ or at $V=0$, we can simplify the above expression even further and write
\be
\langle \Phi (U,V,\phi) \mathcal{O}(t',\phi')\rangle= c \left[U e^{\kappa t'+\frac{r_{-}}{\ell}\Delta \phi} -V e^{-\left(\kappa t'+\frac{r_{-}}{\ell}\Delta \phi \right)}+ \cosh \left( \frac{r_{+}}{\ell} \Delta \phi \right)\right]^{-\Delta} \,.
\ee
Finally, to describe the black hole, we need to take into account the periodicity of the angular variable. This is done by shifting the angular coordinate by integer multiples of $2\pi$ and adding an infinite sum in front of the propagator as\footnote{This procedure is known as the method of the images.} \cite{KeskiVakkuri:1998nw}
\be
\langle \Phi (U,V,\phi) \mathcal{O}(t',\phi')\rangle= c \sum_{n=-\infty}^{n=\infty}\left[U e^{\kappa t'+\frac{r_{-}}{\ell}\Delta \phi_n} -V e^{-\left(\kappa t'+\frac{r_{-}}{\ell}\Delta \phi_n \right)}+ \cosh \left( \frac{r_{+}}{\ell} \Delta \phi_n \right)\right]^{-\Delta}\,,
\ee
where $\Delta \phi_n = \Delta \phi +2 \pi n$.

\subsection{The dual CFT description}
\label{sec: tfd of rotating btz black hole}

The rotating BTZ black hole is dual to a CFT with a chemical potential for angular momentum.
The maximally extended rotating black hole is dual to thermofield double (TFD) state made of two such CFTs 
\bea\label{def: tfd state}
|\text{TFD} \rangle_{t=0} &=& \frac{1}{Z(\beta,\Omega)^{1/2}} \sum_n e^{-\beta \left( E_n +\Omega J_n \right)/2} | E_n ,J_n\rangle_\mt{L} \otimes |E_n, J_n \rangle_\mt{R}\,,\\
Z(\beta,\Omega)&=&\text{Tr} \,e^{-\beta(H-\Omega J)}\,,
\eea
where each CFT has Hamiltonian $H$ and angular momentum $J$, with eigenvalues $E_n$ and $J_n$, respectively. Here $Z(\beta,\Omega)$ is the grand canonical partition function.

We can also decompose the system into left and right moving modes, and write the TFD state as \cite{Caputa:2013eka}
\bea
|\text{TFD} \rangle_{t=0} &=& \frac{1}{Z(\beta_{-},\beta_{+})^{1/2}} \sum_n e^{-\left( \beta_{+} E^\mt{(-)}_{n} +\beta_{-} E^\mt{(+)}_{n} \right)/2} | E^\mt{(-)}_{n} \rangle_\mt{L} \otimes |E^\mt{(+)}_{n} \rangle_\mt{R}\,,\\
Z(\beta,\Omega)&=&\text{Tr} \,e^{-\beta_{-}H_{+}-\beta_{+}H_{-}}\,,
\eea
where the left/right operators $H_{\pm}=\frac{H \pm J}{2}$ have eigenvalues $E^{(\pm)}_{n}$, and the temperatures of the left/right moving modes are
\be
\beta_{\pm}=\beta (1 \mp \ell \Omega)\,.
\ee

\section{Eikonal approximation}
\label{sec: eikonal approximation}

In this section we compute the out-of-time-order correlator (OTOC)
\be \label{OTOC111}
F= \langle\text{TFD}|V_{\phi_1}(t_1) W_{\phi_2}(t_2) V_{\phi_3}(t_3) W_{\phi_4}(t_4) |\text{TFD}\rangle \,,
\ee
using the holographic prescription given be Shenker and Stanford (S\&S) in \cite{Shenker:2014cwa}. Here, to avoid clutter, the subscripts $\phi_1, \phi_2, \cdots$ are used to denote the four angles. We consider all the operators acting on the right side of the geometry\footnote{Two-sided configurations, like, for instance, $\langle V_L W_R V_L W_R \rangle$, can be obtained from the one-sided OTOC by analytic continuation.}. 
For simplicity, we also consider the operators $V$ and $W$ to be single-trace operators, because in this case the corresponding bulk-to-boundary propagators have a simple form.
If the $t_i'$s are all real, the above OTOC diverges. A simple way to regularize this divergence is to consider imaginary times\footnote{Another possibility is smear the operators out in a time scale of order $\beta$.}
\bea
t_1 &=& -t/2+i \epsilon_1\,, \,\,\,\,\,\, t_3 = -t/2+i \epsilon_3\,, \nonumber\\
t_2 &=& t/2+i \epsilon_2\,, \,\,\,\,\,\,\,\,\,\,\, t_4 = t/2+i \epsilon_4\,.
\label{eq-times}
\eea

\begin{figure}[ht!]
\centering
\begin{tikzpicture}[scale=1.]
\draw [thick]  (0,0) -- (0,3);
\draw [thick]  (3,0) -- (3,3);
\draw [thick]  (0,3) -- (1.5,4.5);
\draw [thick]  (1.5,4.5) -- (3,3); 
\draw [thick,dashed]  (0,0) -- (3,3);
\draw [thick,dashed]  (0,3) -- (3,0);
\draw [thick] (0,0) -- (1.5,-1.5);
\draw [thick] (1.5,-1.5) -- (3,0);
\draw [thick,decorate,decoration={zigzag,segment length=1.5mm,amplitude=.5mm}] (0,3) -- (0,4.5);
\draw [thick,decorate,decoration={zigzag,segment length=1.5mm,amplitude=.5mm}] (0,0) -- (0,-1.5);
\draw [thick,decorate,decoration={zigzag,segment length=1.5mm,amplitude=.5mm}] (3,3) -- (3,4.5);
\draw [thick,decorate,decoration={zigzag,segment length=1.5mm,amplitude=.5mm}] (3,0) -- (3,-1.5);



\node[scale=.8,align=center] at (1.0,2.41){$\tilde{U}$};
\node[scale=.8,align=center] at (2.11,2.32){$\tilde{V}$};

\draw[thick,<->] (1,2.2) -- (1.5,1.7) -- (2,2.2);


  \draw [thick,blue] (0.2,0.1) -- (.45,.35);
  \draw [->,thick,blue] (0.65,0.55) -- (.9,.8);
  \draw[-,blue,scale=.15,samples=300,thick,domain=1.5:5.5] plot ({\x},{3+exp(-abs(\x-3.5))*sin(10*(\x-3.5) r)});
  \node[scale=.8,align=center] at (1.1,.8){$p_4^\mt{V}$};

 \draw[-,red,scale=.15,samples=300,thick,domain=15.5:19.5] plot ({\x},{3+exp(-abs(\x-17.5))*sin(10*(\x-17.5) r)});
\draw [thick, red] (2.95,0.15) -- (2.75,0.35);
\draw [->,thick, red] (2.5,0.6) -- (2.25,0.85);
 \node[scale=.8,align=center] at (2.65,.8){$p_3^\mt{U}$};

\node[scale=1,align=center] at (1.5,-1.9){${\color{red}{V_{\phi_3}(t_3)}} {\color{blue}{W_{\phi_4}(t_4)}}| \text{TFD}\rangle$};

\draw [fill=red] (3,.4) circle [radius=0.05];
 \node[scale=1,align=center] at (3.3,.4){$t_3$};

\draw [fill=blue] (3,2.4) circle [radius=0.05];
 \node[scale=1,align=center] at (3.3,2.4){$t_4$};

\draw [thick]  (6,0) -- (6,3);
\draw [thick]  (9,0) -- (9,3);
\draw [thick]  (6,3) -- (7.5,4.5);
\draw [thick]  (7.5,4.5) -- (9,3); 
\draw [thick,dashed]  (6,0) -- (9,3);
\draw [thick,dashed]  (6,3) -- (9,0);
\draw [thick] (6,0) -- (7.5,-1.5);
\draw [thick] (7.5,-1.5) -- (9,0);
\draw [thick,decorate,decoration={zigzag,segment length=1.5mm,amplitude=.5mm}] (6,3) -- (6,4.5);
\draw [thick,decorate,decoration={zigzag,segment length=1.5mm,amplitude=.5mm}] (6,0) -- (6,-1.5);
\draw [thick,decorate,decoration={zigzag,segment length=1.5mm,amplitude=.5mm}] (9,3) -- (9,4.5);
\draw [thick,decorate,decoration={zigzag,segment length=1.5mm,amplitude=.5mm}] (9,0) -- (9,-1.5);

\draw [fill=red] (9,.4) circle [radius=0.05];
 \node[scale=1,align=center] at (9.3,.4){$t_1$};

\draw [fill=blue] (9,2.4) circle [radius=0.05];
 \node[scale=1,align=center] at (9.3,2.4){$t_2$};
 
 \node[scale=1,align=center] at (7.5,-1.9){${\color{blue}{W_{\phi_2}(t_2)^{\dagger}}} {\color{red}{V_{\phi_1}(t_1)^{\dagger}}}| \text{TFD}\rangle$};

 \draw[-,red,scale=.15,samples=300,thick,domain=43:47] plot ({\x},{16+exp(-abs(\x-45))*sin(10*(\x-45) r)});
\draw [->,thick, red] (6.6,2.5) -- (6.3,2.8);
\draw [thick, red] (6.85,2.25) -- (7.05,2.05);

\node [scale=.8,align=center] at (6.5,3.){$p_1^\mt{U}$};

 \draw [thick,blue] (8.4,2.3) -- (8.2,2.1);
  \draw [<-,thick,blue] (8.9,2.8) -- (8.65,2.55);
  \draw[-,blue,scale=.15,samples=300,thick,domain=55:59] plot ({\x},{16+exp(-abs(\x-57))*sin(10*(\x-57) r)});
 \node[scale=.8,align=center] at (8.6,3.0){$p_2^\mt{V}$};

\end{tikzpicture}

\caption{ \small {\it Left}: the `in' state ${\color{red}{V_{\phi_3}(t_3)}} {\color{blue}{W_{\phi_4}(t_4)}}| \text{TFD}\rangle$ represented in a bulk spatial slice that touches the right boundary at time $t_3$. {\it Right}: the `out' state ${\color{blue}{W_{\phi_2}(t_2)^{\dagger}}} {\color{red}{V_{\phi_1}(t_1)^{\dagger}}}| \text{TFD}\rangle$ represented in a bulk slice that touches the right boundary at time $t_2$.}
\label{fig-In-Out-states}
\end{figure}
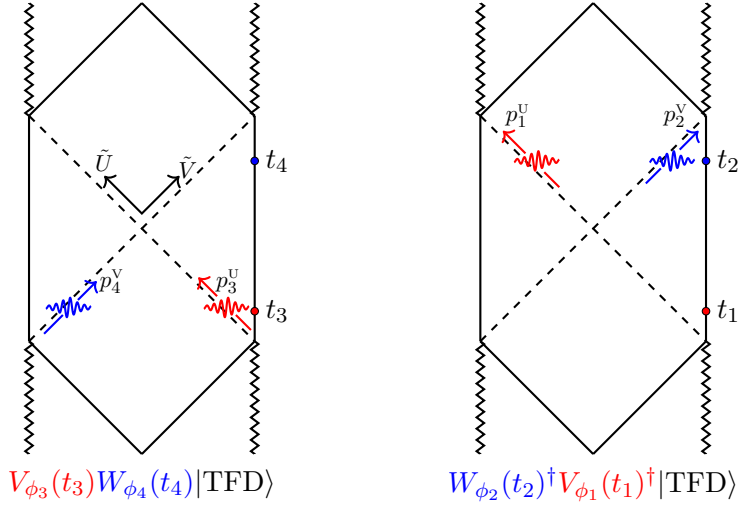

The basic idea of the S\&S approach is to view the OTOC as a scattering amplitude
\be
F = \langle \text{out} |  \text{in} \rangle\,,
\ee
where the `in' and `out' states are given by
\bea
| \text{in} \rangle &=& V_{\phi_3}(t_3) W_{\phi_4}(t_4) |\text{TFD}\rangle\,, \nonumber \\
| \text{out} \rangle &=& W_{\phi_2}(t_2)^{\dagger} V_{\phi_1}(t_1)^{\dagger} |\text{TFD}\rangle\,.
\eea
The above states have a simple description in the bulk in terms of two-particle states. The $V$-operator creates a bulk particle, which we call the $V$-particle, while the particle created by the $W$-operator is called the $W$-particle. For large enough $t$ these particles will be highly boosted with respect to the $t=0$ slice of the geometry. The $W$-particle will have a very large momentum in the $V$-direction, while the $V$-particle will have a very large momentum in the $U$-direction. As explained in \cite{Shenker:2014cwa}, the state $V_{\phi_3}(t_3) W_{\phi_4}(t_4) |\text{TFD}\rangle$ can be thought of as an `in' state, in which the particle's wave functions is represented at some early time (before the collision), while the state $W_{\phi_2}(t_2)^{\dagger} V_{\phi_1}(t_1)^{\dagger} |\text{TFD}\rangle$ can be thought of as an `out' state, with the particle's wave functions represented at some late time (after the collision). See figure \ref{fig-In-Out-states}.

For convenience, we choose to represent the $V$-particle in the $U=0$ slice of the geometry, and we expand the corresponding state in the basis $|p^\mt{U},\phi' \rangle$ of well defined momentum and angular position. Likewise, we represent the $W$-particle in the $V=0$ slice of the geometry, in the basis $|p^\mt{V},\phi' \rangle$. The basis vectors have the following normalization\footnote{This normalization can be fixed by considering the overlap of two Klein-Gordon wave functions. For instance, in the $U=0$ slice of the geometry, the overlap between $\psi$ and $\Phi$ reads $\vev{\psi|\Phi} = (2 i r_{+})\int dV d\phi\, \psi^* \partial_V \Phi$. By writing this overlap in momentum space one can obtain the normalization given in (\ref{eq-norm}).}
\be 
\langle p^\mt{U},\phi' | q^\mt{U},\phi'' \rangle = \frac{A_0^2 p^\mt{U}}{\pi r_{+}} \delta(p^\mt{U}-q^\mt{U}) \delta(\phi'-\phi'')\,,
\label{eq-norm}
\ee
with a similar formula for $| q^\mt{V},\phi' \rangle$. 
Here $A_0$ is defined by $A(0)$ in \eqref{eq-metricBif}.

The `in' state can then be written as
\be
V_{\phi_3}(t_3) W_{\phi_4}(t_4) |\text{TFD}\rangle = \int d\phi_3' \, d\phi_4' \int dp_3^\mt{U} dp_3^\mt{V} \, \psi_3(p_3^\mt{U} ,\phi_3' ) \psi_4(p_4^\mt{V} ,\phi_4' ) |p_3^\mt{U} ,\phi_3' \rangle \otimes | p_4^\mt{V} ,\phi_4' \rangle\,,
\ee
while the `out' state reads
\be
V_{\phi_1}(t_1)^{\dagger} W_{\phi_2}(t_2)^{\dagger} |\text{TFD}\rangle = \int d\phi_1' \, d\phi_2' \int dp_1^\mt{U} dp_2^\mt{V} \, \psi_1(p_1^\mt{U} ,\phi_1' ) \psi_2(p_2^\mt{V} ,\phi_2' ) |p_1^\mt{U} ,\phi_1' \rangle \otimes | p_2^\mt{V} ,\phi_2' \rangle\,.
\ee
 The wave functions $\psi_i$ appearing in the above formulas are Fourier transforms of bulk-to-boundary propagators
\bea
\psi_1(p^\mt{U},\phi') &=& \int  dV e^{i A_0 p^\mt{U} V} \langle \Phi_V(U,V, \phi' ) V_{\phi_1}(t_1)^{\dagger}\rangle|_{U=0} \,, \label{btb1} \\
\psi_2(p^\mt{V},\phi') &=& \int  dU e^{i A_0 p^\mt{V} U} \langle \Phi_W(U,V, \phi' ) W_{\phi_2}(t_2)^{\dagger}\rangle|_{V=0} \,, \label{btb2}\\
\psi_3(p^\mt{U},\phi') &=& \int  dV e^{i A_0 p^\mt{U} V} \langle \Phi_V(U,V, \phi' ) V_{\phi_3}(t_3) \rangle|_{U=0} \,, \label{btb3}\\
\psi_4(p^\mt{V},\phi') &=& \int  dU e^{i A_0 p^\mt{V} U} \langle \Phi_W(U,V, \phi' ) W_{\phi_4}(t_4) \rangle|_{V=0} \,, \label{btb4}
\eea
where $\Phi_V$ and $\Phi_W$ are bulk fields dual to the operators $V$ and $W$, which are represented via the `extrapolate' dictionary.

In the eikonal approximation, the `in' and `out' are related by a phase shift
\be 
\left( |p_1^\mt{U} ,\phi_1' \rangle  \otimes | p_2^\mt{V} ,\phi_2' \rangle \right)_{\text{out}} \approx e^{i \delta(s,\phi_1'-\phi_2')} \left( |p_1^\mt{U} ,\phi_1' \rangle \otimes | p_2^\mt{V} ,\phi_2' \rangle \right)_\text{in}+| \text{inelastic}\rangle\,,
\ee
where $\delta(s,\phi_1'-\phi_2')$ is the so-called {\it Eikonal phase} and $s=(p_1+p_2)^2=2A_0 \,p^\mt{U}p^\mt{V}$ is a Mandelstam variable.
The state $| \text{inelastic}\rangle$ represents an inelastic contribution that is orthogonal to all two-particle `in' states. The physics behind the eikonal approximation is the following: since the particles collide at very high energy, the interaction is dominated by the graviton exchange, and the sum of the corresponding ladder diagrams can be described by a simple phase shift \cite{tHooft:1987vrq,Kabat:1992tb}.

The OTOC \eqref{OTOC111} can then be written as
\be
F= \frac{A_0^4}{\pi^2} \int \int d\phi d\phi' \int \int dp_1^\mt{U} dp_2^\mt{V} e^{i\delta(s,\phi-\phi')} \Big[ p_1^\mt{U} \psi_1^*(p_1^\mt{U},\phi) \psi_3(p_1^\mt{U},\phi) \Big] \Big[p_2^\mt{V} \psi_2^*(p_2^\mt{V},\phi') \psi_4(p_2^\mt{V},\phi') \Big]\,,
\label{eq-OTOCformula}
\ee
where $p_3^\mt{U}=p_1^\mt{U}$, $p_4^\mt{V}=p_2^\mt{V}$,  $\phi_3'=\phi_1' \equiv \phi$ and $\phi_4'=\phi_2'\equiv \phi'$ due to \eqref{eq-norm}.
The information of $\phi_i$ and $t_i$ with $i=1,2,3,4$ in \eqref{OTOC111} is denoted by the subscripts of $\psi_i$.

\subsection{Wave functions}

For the rotating BTZ black hole the bulk-to-boundary propagators in \eqref{btb1}and \eqref{btb2} are given by
\bea
& &\langle \Phi_V(0,V,\phi) V_{\phi_1}(t_1)^\dagger\rangle = c_\mt{V} \sum_{n=-\infty}^{n=\infty}\left[ -V e^{-\left(\kappa t_1^*+\frac{r_{-}}{\ell}\Delta \phi_n \right)}+ \cosh \left( \frac{r_{+}}{\ell} \Delta \phi_n \right)\right]^{-\Delta_V}\,,\\
& &\langle  \Phi_W(U,0,\phi') W_{\phi_2}(t_2)^\dagger\rangle = c_\mt{W} \sum_{n=-\infty}^{n=\infty}\left[ U e^{\kappa t_2^*+\frac{r_{-}}{\ell}\Delta \phi'_n }+ \cosh \left( \frac{r_{+}}{\ell} \Delta \phi'_n \right)\right]^{-\Delta_W}\,,
\eea
where 
\be
\Delta \phi_{n} = \phi -\phi_1 +2\pi n\,, \,\,\,\,\,\, \Delta \phi'_{n} = \phi' -\phi_2 +2\pi n\,.
\ee
For the bulk-to-boundary propagators in \eqref{btb3}and \eqref{btb4} we only need to change $t \rightarrow t^*$ with 
\be
\Delta \phi_{n} = \phi -\phi_3 +2\pi n\,, \,\,\,\,\,\, \Delta \phi'_{n} = \phi' -\phi_4 +2\pi n\,.
\ee
To avoid clutter, we use the same notation $\Delta \phi_n$ for $\phi_1$ and $\phi_3$ ($\Delta \phi_n'$ for $\phi_2$ and $\phi_4$).

The wave functions can then be written as
\bea
\psi_1(p^\mt{U},\phi)&=&-\theta(p^\mt{U}) \frac{2\pi i c_V}{\Gamma(\Delta_V)}  \sum_{n=-\infty}^{n=+\infty}  e^{\kappa t_1^\mt{*}+\frac{r_{-}}{\ell} \Delta \phi_n} \left(-i A_0 p^\mt{U} e^{\kappa t_1^\mt{*}+\frac{r_{-}}{\ell} \Delta \phi_n} \right)^{\Delta_V-1} e^{i A_0 p^\mt{U} \cosh\left(\frac{r_{+}}{\ell} \Delta \phi_n \right) e^{\kappa t_1^\mt{*}+\frac{r_{-}}{\ell} \Delta \phi_n} }\,, \nonumber \\
\psi_3(p^\mt{U},\phi)&=&-\theta(p^\mt{U}) \frac{2\pi i c_V}{\Gamma(\Delta_V)}  \sum_{n=-\infty}^{n=+\infty}  e^{\kappa t_3+\frac{r_{-}}{\ell} \Delta \phi_n} \left(-i A_0 p^\mt{U} e^{\kappa t_3+\frac{r_{-}}{\ell} \Delta \phi_n} \right)^{\Delta_V-1} e^{i A_0 p^\mt{U} \cosh\left(\frac{r_{+}}{\ell} \Delta \phi_n \right) e^{\kappa t_3+\frac{r_{-}}{\ell} \Delta \phi_n} }\,, \nonumber\\
\psi_2(p^\mt{V},\phi')&=&\theta(p^\mt{V}) \frac{2\pi i c_W}{\Gamma(\Delta_W)}  \sum_{n=-\infty}^{n=+\infty}  e^{-\kappa t_2^\mt{*}-\frac{r_{-}}{\ell} \Delta \phi'_n} \left(i A_0 p^\mt{V} e^{\kappa t_2^\mt{*}+\frac{r_{-}}{\ell} \Delta \phi'_n} \right)^{\Delta_W-1} e^{-i A_0 p^\mt{V} \cosh\left(\frac{r_{+}}{\ell} \Delta \phi'_n \right) e^{-\kappa t_2^\mt{*}-\frac{r_{-}}{\ell} \Delta \phi'_n} }\,, \nonumber\\
\psi_4(p^\mt{V},\phi')&=&\theta(p^\mt{V}) \frac{2\pi i c_W}{\Gamma(\Delta_W)}  \sum_{n=-\infty}^{n=+\infty}  e^{-\kappa t_4-\frac{r_{-}}{\ell} \Delta \phi'_n} \left(i A_0 p^\mt{V} e^{\kappa t_4+\frac{r_{-}}{\ell} \Delta \phi'_n} \right)^{\Delta_W-1} e^{-i A_0 p^\mt{V} \cosh\left(\frac{r_{+}}{\ell} \Delta \phi'_n \right) e^{-\kappa t_4-\frac{r_{-}}{\ell} \Delta \phi'_n} }\,. \nonumber \\
\,\,
\eea

\subsection{The eikonal phase}

The operators $V$ and $W$ introduce particles in the bulk, which we call the $V$-particle and the $W$-particle, respectively. In the elastic eikonal gravity approximation, the phase shift is basically given by the on-shell action of these particles, namely
\be
\delta(s,\phi-\phi') = S_\text{classical}\,,
\ee
where $S_\text{classical}$ is simply the sum of the on-shell actions for the $V$-particles and $W$-particles. When the $W$ is very late, and the $V$ operator is very early, the corresponding bulk particles follow an almost null trajectory very close the black hole horizon. The $W$-particle moves along $U=0$, while the $V$-particle moves along $V=0$. 
The particles collide very close to the bifurcation surface, at $U=V=0$, and the phase shift is only sensitive to this region of the geometry, where the metric assumes the approximate form
\be 
ds^2=-2A(UV) dU dV+r^2(UV) d\phi^2\,.
\label{eq-metricBif}
\ee
In the following, we denote the horizon values by $A_0 \equiv A(0)$ and $r_+ \equiv r(0)$. In this particular configuration, the stress-energy of these particles takes the form
\bea
&&W\text{-particle}: T_{UU}=\frac{A_0}{r_{+}}p_2^\mt{V} \delta(U) \delta(\phi-\phi'')\,,\nonumber\\
&&V\text{-particle}: T_{VV}=\frac{A_0}{r_{+}}p_1^\mt{U} \delta(V) \delta(\phi-\phi')\,.
\eea
where $\phi'$ and $\phi''$ denote the position of the sources.
The back-reaction of these particles in the geometry takes a simple shock wave form. For instance, the backreaction of the $W$-particles reads
\be
ds^2 \rightarrow ds^2 + h_{UU} dU^2\,,\,\,\,\, h_{UU}= 16 \pi G_N r_{+} A_0 p_2^\mt{V} \delta(U) f(\phi-\phi'') \,,
\ee
while the back-reaction of the $V$-particle reads 
\be
ds^2 \rightarrow ds^2 + h_{VV} dV^2\,,\,\,\,\, h_{VV}= 16 \pi G_N r_{+} A_0 p_1^\mt{U} \delta(V) f(\phi-\phi') \,.
\ee
In the above formulas, $f(\phi)$ is the shock wave transverse profile. For a rotating BTZ black hole, $f(\phi)$ is a solution of the equation\footnote{This equation is obtained from the linearized Einstein's equations. As a result, our shock wave solutions are only valid at the perturbative level. However, the phase shift is only sensitive to the region very close to the bifurcation surface, where the above shock wave solutions are exact solutions of Einstein's equations. }
\be
\ell^2 f''(\phi)-2\ell r_{-}f'(\phi)-(r_{+}^2-r_{-}^2)f(\phi)=\delta(\phi) \,.
\ee
The most general solution of the above equation has the form
\be
f(\phi)=c_1 e^{(r_{+}+ r_{-})\phi/\ell}+c_2 e^{-(r_{+}-r_{-})\phi/\ell} \,,
\label{eq-f}
\ee
where $c_1$ and $c_2$ are constants and we assumed $\phi>0$\footnote{For a more detailed discussion of this solution and the constants, see \cite{Fu:2018oaq}. Notice, however, that our co-rotating coordinate differ from the one used in that reference by a minus sign, i.e., $\phi_\mt{here}=-\phi_\mt{there}$.}.

The on-shell classical action for the $W$-particles and $V$-particles can be written as \cite{Shenker:2014cwa}
\be
S_\text{classical} =\frac{1}{2}\int d^3x \sqrt{-g}\, h_{UU}T^{UU} \,.
\ee
Note that, while $h_{UU}$ refers to the $W-$particle, the stress-tensor $T^{UU}=g^{UV}g^{UV}T_{VV}$ refers to the $V-$particle. Substituting the expression for the stress-energy tensors and the corresponding back-reactions, one finds
\be
\label{eq-phaseshift}
\delta(s,\phi'-\phi'')= 4\pi G_N r_{+} s\,f(\phi'-\phi'')\,,\,\,\,\,s=2A_0 \,p_1^\mt{U} p_2^\mt{V}\,.
\ee
\subsection{Evaluating the integrals}

Inserting the above wave functions in the formula for the OTOC (\ref{eq-OTOCformula}), we find
\begin{multline}
F=\left( \frac{4\pi c_V c_W}{\Gamma(\Delta_V) \Gamma(\Delta_W)} \right)^2 A_0^{2(\Delta_V+\Delta_W)} \sum_{n_1, n_2,n_3, n_4} \int d\phi d\phi' dp^\mt{U} dp^\mt{V} e^{i \delta(s,\phi-\phi')} (p^\mt{U})^{2\Delta_V-1} (p^\mt{V})^{2\Delta_W-1} \times \\
\times \frac{e^{ \Delta_V \kappa (t_1+t_3)}}{e^{ \Delta_W \kappa (t_2+t_4)}}  \frac{e^{\Delta_V \frac{r_{-}}{\ell} (\Delta \phi_{n_1}+\Delta \phi_{n_3})}}{e^{\Delta_W \frac{r_{-}}{\ell} (\Delta \phi'_{n_2}+\Delta \phi'_{n_4})}} \frac{e^{i A_0 p^\mt{U}\left[ \cosh\left(\frac{r_+}{\ell} \Delta \phi_{n_3} \right)e^{\kappa t_3+\frac{r_-}{\ell} \Delta \phi_{n_3}} - \cosh\left(\frac{r_+}{\ell} \Delta \phi_{n_1} \right)e^{\kappa t_1+\frac{r_-}{\ell} \Delta \phi_{n_1}} \right] }}{ e^{i A_0 p^\mt{V}\left[ \cosh\left(\frac{r_+}{\ell} \Delta \phi'_{n_4} \right)e^{\kappa t_4+\frac{r_-}{\ell} \Delta \phi'_{n_4}} - \cosh\left(\frac{r_+}{\ell} \Delta \phi'_{n_2} \right)e^{\kappa t_2+\frac{r_-}{\ell} \Delta \phi'_{n_2}} \right] }}\,,
\end{multline}
where $\Delta \phi_{n_i}=\phi-\phi_1+2\pi n_i$ and $\Delta \phi'_{n_i}=\phi'-\phi_2+2\pi n_i$.

The OTOC can be divided into two contributions
\be
F=F_{0}+F_{n_1,n_2,n_3,n_4} \,,
\ee
where $F_0$ corresponds to the term where $n_1=n_2=n_3=n_4=0$, and $F_{n_1,n_2,n_3,n_4}$ corresponds to the remainder terms. The dominant contribution comes from $F_0$
\begin{multline}
F_0=\left( \frac{4\pi c_V c_W}{\Gamma(\Delta_V) \Gamma(\Delta_W)} \right)^2 A_0^{2(\Delta_V+\Delta_W)} \int d\phi d\phi' dp^\mt{U} dp^\mt{V} e^{i\delta(s,\phi-\phi')} (p^\mt{U})^{2\Delta_V-1} (p^\mt{V})^{2\Delta_W-1} \times \\
\times \frac{e^{ \Delta_V \kappa (t_1+t_3)}}{e^{ \Delta_W \kappa (t_2+t_4)}}  \frac{e^{2\Delta_V \frac{r_{-}}{\ell} \Delta \phi}}{e^{2\Delta_W \frac{r_{-}}{\ell} \Delta \phi'}} \frac{e^{i A_0 p^\mt{U} \cosh\left(\frac{r_+}{\ell} \Delta \phi \right)e^{\frac{r_-}{\ell} \Delta \phi} \left( e^{\kappa t_3 } - e^{\kappa t_1} \right) }}{ e^{i A_0 p^\mt{V} \cosh\left(\frac{r_+}{\ell} \Delta \phi' \right)e^{\frac{r_-}{\ell} \Delta \phi'} \left( e^{\kappa t_4 } - e^{\kappa t_2} \right) }} \,.
\end{multline}
The only ingredient missing to calculate $F_0$ is the phase shift $\delta(s,\phi-\phi')$, which we know to have the following form (see (\ref{eq-phaseshift}))
\be
\delta(s,\phi-\phi') = 8\pi r_+ G_N p^\mt{U} p^\mt{V} f(\phi-\phi')\,,
\ee
where $f(\phi-\phi')$ is given by (\ref{eq-f}).

To evaluate the above integral it is convenient to introduce the new variables
\bea
p&=& -i A_0 p^\mt{U} \left( e^{\kappa t_3}-e^{\kappa t_1}\right)\,,\nonumber \\
q&=& i A_0 p^\mt{V} \left( e^{\kappa t_4}-e^{\kappa t_2}\right)\,.
\eea
By specifying the times as in (\ref{eq-times}), the integral becomes
\begin{multline}
F_0=C \int d\phi d\phi' dp \,dq \,p^{2\Delta_V-1} q^{2\Delta_W-1} e^{i 8\pi r_{+} G_N A_0^{-1} \frac{p q }{ \epsilon_{13} \epsilon^\mt{*}_{24}}e^{\kappa t}f(\phi-\phi')} \times \\ 
\times e^{ \left[ \frac{2r_-}{\ell} \left( \Delta_V \Delta \phi- \Delta_W \Delta \phi' \right)-p \cosh\left( \frac{r_+}{\ell} \Delta \phi \right)e^{\frac{r_-}{\ell}\Delta \phi}-q \cosh\left( \frac{r_+}{\ell} \Delta \phi' \right)e^{\frac{r_-}{\ell}\Delta \phi'} \right]} \,,
\end{multline}
where $\epsilon_{ij}=i(e^{\kappa \epsilon_i}-e^{\kappa \epsilon_j})$ and $C$ is a constant given by
\be
C=\frac{2 \pi^2 c_V^2 c_W^2}{\Gamma(\Delta_V)^2 \Gamma(\Delta_W)^2} \left[ \frac{1}{2 \sin\left( \frac{\epsilon_3-\epsilon_1}{2}\right)}\right]^{2\Delta_V} \left[ \frac{1}{2 \sin\left( \frac{\epsilon_4-\epsilon_2}{2}\right)}\right]^{2\Delta_W}\,.
\ee

Without the phase shift, the above integral just gives $\langle V V\rangle \langle W W \rangle$. The integral can be evaluated in the limit $\Delta_W >> \Delta_V >>1$ and the result reads\footnote{Here we evaluate the integrals using the same tricks that were used in \cite{Shenker:2014cwa}. By writing the integrals in $q$ and $\phi'$ as $\int dq d\phi' e^{-F(q,\phi')}$, we can check that the result is dominated by the region where $q\approx 2\Delta_W$ and $\phi' \approx \phi_2$. The integral in $p$ can be done analytically, and the integral in $\phi$ can be done by a saddle point approximation.}
\bea
\text{OTOC}(t,\phi_{12})&=&\frac{\langle V_{\phi_1}(t_1) W_{\phi_2}(t_2) V_{\phi_1}(t_3) W_{\phi_2}(t_4) \rangle }{\langle V_{\phi_1}(i \epsilon_1) V_{\phi_1}(i \epsilon_3)\rangle \langle W_{\phi_1}(i \epsilon_2) W_{\phi_4}(i \epsilon_4) \rangle}=\frac{1}{\left[ 1+ \frac{i 16 \pi G_N \Delta_{W}}{ \ell^2 \epsilon_{13} \epsilon^\mt{*}_{24}}e^{\kappa t} f(\phi_{12})\right]^{\Delta_V}} \nonumber \\
&\approx & 1+\frac{i 16 \pi G_N \Delta_V \Delta_{W}}{ \ell^2 \epsilon_{13} \epsilon^\mt{*}_{24}} e^{\kappa t} f(\phi_{12}) \,,
\eea
where $\phi_{12}\equiv\phi_1-\phi_2$. Replacing $f(\phi)$ by (\ref{eq-f}) and using that
\be
\frac{r_{+}+r_{-}}{\ell}=\frac{2\pi}{\beta}\frac{\ell}{1-\ell\, \Omega}\,,\,\,\,\, \frac{r_{+}-r_{-}}{\ell}=\frac{2\pi}{\beta}\frac{\ell}{1+\ell\, \Omega}\,,
\ee
we can write the OTOC as 
\be \label{main11}
\text{OTOC}(t,\phi_{12}) \approx 1+ C_1 e^{\frac{2\pi}{\beta}\left(t+\frac{\ell \phi_{12}}{1-\ell \, \Omega} \right)}+C_2 e^{\frac{2\pi}{\beta}\left(t-\frac{\ell \phi_{12}}{1+\ell \, \Omega} \right)} \,,
\ee
where the constants $C_1$ and $C_2$ are proportional to $\Delta_V \Delta_W G_N$.
These two solutions correspond to left and right moving modes. Both modes have a maximal Lyapunov exponent, $\lambda_L=\frac{2\pi}{\beta}$, but they have different butterfly velocities
\be
\frac{v_B^{\pm}}{c}=1\pm \ell \Omega \,.
\ee
One of these butterfly velocities is larger than the speed of light when $\Omega >0$. 

In terms of the Schwarzschild angular coordinate, $\varphi_{12}=\phi_{12}+\Omega \,t_{12}$ (here $t_{12}=\text{Re}\,(t_1-t_2)=-t$), we have
\be 
\text{OTOC}(t,\varphi_{12}) \approx 1+ C_1 e^{\frac{2\pi}{\beta_{+}}\left(t+\ell \varphi_{12}\right)}+C_2 e^{\frac{2\pi}{\beta_{-}}\left(t-\ell \varphi_{12} \right)}\,,
\ee
where
\be
\beta_{\pm}=\beta (1\mp \ell\,\Omega)\,.
\ee
In this case the two butterfly velocities are equal to the speed of light, but the two modes have different Lyapunov exponents
\be
\lambda^{\pm}_{L}=\frac{2\pi}{\beta_{\pm}}\,,
\ee
which satisfy the property $\lambda^{-}_L \leq \lambda_L \leq \lambda^{+}_L$. Naively, this seems to indicate that one of the Lyapunov exponents is less than maximum, while the other one violates the chaos bound \cite{Maldacena:2015waa}. We come back to this issue in the discussion section.

\section{Chaos in the Chern-Simons formulation of AdS$_3$ gravity}
\label{sec: cs gravity review}

In the previous section, we calculated the Lyapunov exponent and the butterfly velocity using the gravity eikonal approximation. In this section, we use the $SL(2,\mathbb{C})$ Chern-Simons formulation of Euclidean $AdS_3$ gravity to derive the on-shell action of the boundary graviton. This is analogous to the Schwarzian on-shell action of the $2D$ Jackiw-Teitelboim model~\cite{Maldacena:2016upp,Jensen:2016pah,Bagrets:2016cdf,Grumiller:2017qao,Mandal:2017thl} as well as the dimensional reduction from higher dimensional black holes~\cite{Cvetic:2016eiv,Das:2017pif,Das:2017hrt,Castro:2018ffi,Gaikwad:2018dfc}. Moreover, by using a metric-like formulation, \cite{Poojary:2018esz} obtained a similar on-shell action, which is equivalent to our result up to quadratic level. Since we will evaluate the leading Lyapunov exponent in large $\kc\sim c$ from the quadratic on-shell action, the leading Lyapunov exponent agrees with~\cite{Poojary:2018esz}. Also, the on-shell action from the Chern-Simons gravity and Lyapunov exponent thereof can easily be generalized for the case of higher spin gravity~\cite{Narayan:2019}.

\subsection{Review of Chern-Simons formulation of AdS$_3$ gravity}


In this section, we review the $SL(2,\mathbb{C})$ Chern-Simons formulation of Euclidean AdS$_3$ (EAdS$_3$) gravity. In particular, we follow~\cite{deBoer:2013gz} in the choice of the boundary term and the corresponding boundary condition. The case of Lorentzian $AdS_3$ is similar, except that one has to choose carefully the ``Kruskal-like'' gauges for the eternal black hole~\cite{Castro:2016ehj}. 

The Euclidean solution can be obtained by analytic continuation of \eqref{eq-metricS} to imaginary values of $t$ and $r_{-}$
\be
t=-it_E\,,\,\,\,\,\,\,r_{-}=i r_E\,.
\ee
The Euclidean coordinates $(t_E,\varphi)$ satisfy the periodic condition~\cite{Carlip:1994gc,KeskiVakkuri:1998nw} 
\be
(t_E,\varphi) \sim (t_E + \beta,\varphi+w)\,,
\ee
where
\be
w \equiv -i \beta\, \Omega= \beta{r_E\over r_+ l }\,.
\ee
Note that $w$ and $r_E$ are taken to be real in the Euclidean black hole, but they become purely imaginary in the Lorentzian BTZ black hole.
%
We define holomorphic and anti-holomorphic coordinates $(z,\bar{z})$ as
\be
z \equiv \varphi+i\frac{t_E}{\ell}\,,\,\,\,\,\,\,\bar{z}\equiv\varphi-i\frac{t_E}{\ell}\,.
\ee
The modular parameters $\tau$ and $\bar{\tau}$ are defined as
\be
	\tau \equiv w + {i\beta \over l}\,,\,\,\,\,\,\,\bar{\tau} \equiv w - {i\beta \over l}\,.
\ee
In terms of the $(z,\bar{z})$ and $(\tau,\bar{\tau})$, the above periodic condition becomes
\begin{equation}
	(z,\bar{z})\sim (z+\tau,\bar{z}+\bar{\tau})\,.
\end{equation}

In three dimensions, EAdS$_3$ gravity can be described by the Chern-Simons action with complex Lie algebra $SL(2,\mathbb{C})$~\cite{Achucarro:1987vz,Witten:1988hc,Krasnov:2001cu,Arcioni:2002vv}
\begin{equation}
	I_{CS}={i \kc\over 4\pi }\int_M \tr \left[ CS(A)-CS(\bar{A})\right]\,,\label{def: cs action}
\end{equation}
where $\kc={l\over 4G}$ is Chern-Simons level and $CS(A)$ is defined by
\begin{equation}
	   CS(A)= A\wedge dA +{2\over 3} A\wedge A \wedge A\,,
\end{equation}
Here, $A$ is the Chern-Simons connection, and $\bar{A}$ is its conjugate, defined by
\begin{equation}
	\bar{A}=-A^\dag\,.
\end{equation}
We use the coordinates $(r,z,\bar{z})$ where $z$ and $\bar{z}$ are defined by
\begin{equation}
    z\equiv\varphi+ i {t_E\over l}\hspace{3mm},\hspace{5mm} \bar{z}\equiv \varphi- i {t_E\over l}\,.
\end{equation}
For EAdS$_3$ Chern-Simons gravity, the manifold in our consideration is a solid torus, and the modular parameter $\tau$ of the boundary torus gives the periodicity of $z$ coordinate
\begin{equation}
    z\sim z+2\pi \sim z+\tau\,.
\end{equation}
It is useful to fix the gauge as~\cite{Krasnov:2001cu,Arcioni:2002vv,Campoleoni:2010zq,Campoleoni:2011hg,deBoer:2013gz} 
\begin{align}
    A=&b^{-1} (d+ a_z dz + a_{\bar{z}}d\bar{z} ) b \,, \\
	\bar{A}=&b (d+ \bar{a}_z dz + \bar{a}_{\bar{z}}d\bar{z}) b^{-1} \,,
\end{align}
where $b(r)$ is defined by
\begin{equation}
	b=e^{rL_0}\,.
\end{equation}
Here, we define the $sl(2)$ generators $L_0,L_{\pm1}$ as
\begin{equation}
    L_0=\begin{pmatrix}
    {1\over 2} & 0\\
    0 & -{1\over 2}\\
    \end{pmatrix}\,,\hspace{3mm}L_1=\begin{pmatrix}
    0 & 0\\
    1 & 0\\
    \end{pmatrix}\,,\hspace{3mm}L_{-1}=\begin{pmatrix}
    0 & -1 \\
    0 & 0\\
    \end{pmatrix}\,.
\end{equation}
The asymptotic AdS$_3$ solution with a flat boundary metric in the Chern-Simons gravity was found to be~\cite{Campoleoni:2010zq,Campoleoni:2011hg,deBoer:2013gz}
\begin{align}
    \left.A-A_{\ads}\right|_{\partial\mathcal{M}}\sim \mathcal{O}(1)\,,\label{eq: asymptotic ads condition}
\end{align}
where $A_{\ads}$ is the $AdS_3$ exact solution
\begin{equation}
    A_{\ads}=b^{-1}\left(L_1 + {1\over 4}L_{-1}\right) b dz + b^{-1} db \,,\hspace{6mm} \bar{A}_{\ads}=b\left(L_{-1} + {1\over 4}L_1\right) b^{-1}d\bar{z}+ b db^{-1}\,.
\end{equation}
Hence, the asymptotic AdS$_3$ condition implies that
\begin{equation}
    a=L_1dz +\cdots\,,\hspace{5mm} \bar{a}=L_{-1}d\bar{z} +\cdots \,.
\end{equation}
The variation of the action~\eqref{def: cs action} without the additional boundary term is given by
\begin{equation}
    \delta I_{CS}= -{i\kc \over 4\pi } \int_{\partial\mathcal{M} } \tr \left[ A \wedge \delta A- \bar{A} \wedge \delta \bar{A} \right]\,.
\end{equation}
Hence, choosing the boundary condition~$A_{\bar{z}}=\bar{A}_z=0$ together with the gauge symmetry~\cite{Campoleoni:2010zq,Campoleoni:2011hg}, we can fix the gauge as follows
\begin{align}
    a=& \left(L_1 - {2\pi \over \kc} \mathcal{L}(z)L_{-1} \right)dz\,,\\
    \bar{a}=& \left(L_{-1} - {2\pi \over \kc} \bar{\mathcal{L}}(\bar{z} ) L_{1} \right)d\bar{z}\,.
\end{align}

Now, we have to take the variation of the modular parameter $\tau$ because it can also be varied under the variation of the action. For this, we will summarize the $(\tau,\bar{\tau})$ formalism discussed in~\cite{deBoer:2013gz}. It is useful to introduce the new coordinates $(w,\bar{w})$, defined by
\begin{equation}
	z={1-i {\tau\over 2\pi} \over 2} w + {1+ i {\tau\over 2\pi}\over 2 } \bar{w}\,,
\end{equation}
which has a fixed periodicity
\begin{equation}
    w\sim w+2\pi \sim w+2\pi i\,.
\end{equation}
At a cost of fixing periodicity, the modular parameter appears in the boundary volume element as well as in the boundary metric
\begin{align}
	ds^2=& \, dzd\bar{z} = \left|\left({1-i{\tau\over 2\pi}\over 2}\right)dw +\left({1+i{\tau\over 2\pi} \over 2 }\right)d\bar{w} \right|^2\,,\\
	i dw  \wedge &d\bar{w}= {4\pi dz^2 \over \im (\tau)}\,,
\end{align}
where $dz^2={i\over 2} dz\wedge d\bar{z}$. The key idea of~\cite{deBoer:2013gz} is to keep the (boundary) volume element of $(w,\bar{w})$ fixed under the variation. In $(w,\bar{w})$ coordinates, the differential form is not varied under the variation of the bulk action in~\eqref{def: cs action}. Instead, in returning to $(z,\bar{z})$ coordinates, the variation of $a_w, a_{\bar{w}}$ gives the variation of the modular parameter $\tau,\bar{\tau}$ because it appears in the transformation of $(a_w,a_{\bar{w}})$ into $(a_z,a_{\bar{z}})$
\begin{equation}
    a_w=\left({1-i{\tau\over 2\pi} \over 2 }\right) a_z +\left({1-i{\bar{\tau}\over 2\pi} \over 2 }\right) a_{\bar{z}}\,.
\end{equation}
For the variational principle, the authors of \cite{deBoer:2013gz} choose the boundary term given by
\begin{equation}
	I_b=-{\kc \over 2\pi } \int_{\partial \mathcal{M} } d^2 z\; \tr \left(  \left(a_z- 2 L_1 \right)a_{\bar{z}} \right) -{\kc \over 2\pi } \int_{\partial \mathcal{M}} d^2 z\; \tr \left(  (\bar{a}_{\bar{z}}- 2 L_{-1})\bar{a}_{z} \right) \,,
\end{equation}
and the variation of the total action $I_{\text{\tiny tot}}\equiv I_{CS}+I_b$ is found to be
\begin{align}
	\delta I_{\text{\tiny tot}} =&- i \kc \int_{\partial \mathcal{M} } {d^2z\over 2\pi \im (\tau) } \tr \left[ (a_z-L_1)\delta\left( (\bar{\tau}-\tau)a_{\bar{z}}\right)+\left({a_z^2\over 2}+ a_za_{\bar{z}} -{\bar{a}_z^2\over 2} \right)\delta \tau \right.\cr
	&\hspace{35mm}\left.-(-\bar{a}_{\bar{z}}-L_{-1})\delta\left( (\bar{\tau}-\tau)\bar{a}_{z}\right)+\left({\bar{a}_{\bar{z}}^2\over 2}+ \bar{a}_{\bar{z}}\bar{a}_{z} -{a_{\bar{z}}^2\over 2} \right)\delta \bar{\tau} \right]\,.
\end{align}
Hence, we can impose the boundary conditions 
\begin{align}
	a_{\bar{z}}= \bar{a}_z=0\,,\qquad \delta\tau=\delta\bar{\tau}=0\,.\label{eq: boundary condition}
\end{align}
With these boundary conditions, the on-shell action of $I_{\text{\tiny tot}}$ is found to be
\begin{align}
	I_\ons =  {i \kc\over 2\pi} \int {d^2 z\over \im (\tau) }  \tr\left[ {\tau\over 2}a_z^2  - {\bar{\tau}\over 2}\bar{a}^2_{\bar{z}}   \right]\,.\label{eq: onshell action deboer}
\end{align}
Note that though the authors of \cite{deBoer:2013gz} mainly analyzed a constant solution $a$ and $\bar{a}$, they pointed out that their method could be applied to non-constant solutions. Furthermore, the $sl(2,\mathbb{C})$ Chern-Simons gravity is simple because an additional chemical potential is not necessary, as opposed to the case of higher spin gravity. In the next section, we will analyze the non-constant solutions and their on-shell action.

\subsection{On-shell action of asymptotic AdS solutions}
\label{sec: on-shell action}

As in the previous section, the boundary condition in \eqref{eq: boundary condition} allows us to fix the connections to be
\begin{equation}
	a_z(z)=\begin{pmatrix}
	0 & { 2\pi  \over \kc } \mathcal{L}(z) \\
	1 & 0 \\
	\end{pmatrix}\,, \qquad 
	\bar{a}_{\bar{z}}(\bar{z}) =\begin{pmatrix}
	0 & -1 \\
	-{ 2\pi  \over \kc } \bar{\mathcal{L}}(\bar{z})  & 0 \\
	\end{pmatrix}\,.\label{eq: gauge condition}
\end{equation}
Then, the on-shell action in \eqref{eq: onshell action deboer} can be written as
%
%
\begin{align}
	I_\ons=  i \int {d^2 z\over \im (\tau) }  \left[ \tau \mathcal{L}(z) - \bar{\tau}  \bar{\mathcal{L}}(\bar{z}) \right]\,.\label{eq: onshell action}
\end{align}
We consider non-constant connections, which are smoothly connected to a fixed constant solution, such as the BTZ black hole, \ie
\begin{equation}
	a= h^{-1} a_{\const}h+ h^{-1}d h \,,
\end{equation}
where $h=h(z)$ is a holomorphic residual gauge transformation parameter that can be smoothly connected to identity, and the constant solution $a_{\const}$ is given by
\begin{equation}
	a_\const=\begin{pmatrix}
	0 & {2\pi \over \kc}\mathcal{L}_0\\
	1 & 0 \\
	\end{pmatrix}\,,
\end{equation}
with $\mathcal{L}_0$ being a constant. Such a smooth residual gauge transformation does not change the holonomy (up to a similarity transformation)
\begin{equation}
	\text{Hol}_{\mathcal{C}}(A)\equiv\mathcal{P}\exp \left[-\int_{\mathcal{C}} A \right]=b^{-1}h^{-1} e^{-\mathbb{w}} h b \,.
\end{equation}
The smoothness of a holonomy along the contractible cycle of solid torus requires the holonomy to belong to the center of the gauge group~\cite{Gutperle:2011kf,Castro:2011fm,Castro:2011iw}. For the BTZ black hole, the Euclidean time circle corresponds to the contractible cycle, and the smoothness condition implies that
\begin{equation}
	\mathbb{w}=\tau a_z +\bar{\tau} a_{\bar{z}}=u^{-1} ( 2\pi i L_0) u \,,
\end{equation}
for some matrix $u$. Hence, by taking the determinant of the both sides
\begin{align}
	\det (\mathbb{w})=& \det \begin{pmatrix}
	0 & { 2\pi \tau \mathcal{L}_0 \over \kc }  \\
	\tau   & 0 \\
	\end{pmatrix}=-{ 2\pi\tau^2  \mathcal{L}_0 \over \kc }  = \pi^2\,,
\end{align}
we obtain
\begin{align}
	\tau =i\pi \sqrt{\kc \over 2\pi \mathcal{L}_0 }\hspace{3mm},\hspace{8mm} \bar{\tau} = -i \pi \sqrt{\kc \over 2\pi \bar{\mathcal{L}}_0 }\,.
\end{align}
In the variation of the action, we consider a fixed constant modular parameter $\tau$ and $\bar{\tau}$, and, accordingly, the smoothness condition of the holonomy also fixes the constant solution $\mathcal{L}_0$ and $\bar{\mathcal{L}}_0$. This means that we are considering the BTZ black hole with $\tau,\bar{\tau}$ and its smooth fluctuations.

Now, we evaluate the non-constant connection $a_z(z)$, which is connected to the constant connection $a_\const$ via a smooth residual gauge transformation by $h(z)$. In principle, using the Gauss decomposition of $h(z)$, one can find the residual gauge symmetry parameter~\cite{Coussaert:1995zp}. However, within this method, it is not clear how to distinguish the large gauge transformation from the smooth one (see Appendix~\ref{app: finite transf}) 

Alternatively, one may consider an infinitesimal gauge transformation, which guarantees the smoothness of gauge transformation. For example, let us consider the following (infinitesimal) gauge transformation
\begin{equation}
    h(z)=\begin{pmatrix}
    1 & 0 \\
    0 & 1\\
    \end{pmatrix}+ \lambda(z)+\cdots\,, \hspace{8mm}\mbox{where} \qquad \lambda(z)\equiv \begin{pmatrix}
    {1\over 2}\lambda_0(z) & -\lambda_{-1}(z) \\
    \lambda_1(z) & -{1\over 2}\lambda_0(z)\\
    \end{pmatrix}\,.
\end{equation}
Under the infinitesimal gauge transformation from $a_\const$
\begin{equation}
    \delta a = \partial_z \lambda(z) + [a_\const, \lambda(z)] \,,
\end{equation}
one can express $\lambda_0(z)$ and $\lambda_{-1}(z)$ in terms of $\lambda_1(z)$ by demanding the transformation to keep the gauge condition in~\eqref{eq: gauge condition}. The residual gauge transformation parametrized by $\lambda_1(z)$ leads to the transformation of the constant $\mathcal{L}_0$:
\begin{equation}
    \delta \mathcal{L} = 2 \partial_z \lambda_1 \mathcal{L}_0 -{\kc \over 4\pi }\partial_z^3 \lambda_1\,.
\end{equation}
This is a special case of the asymptotic symmetry of AdS$_3$~\cite{Campoleoni:2010zq,Campoleoni:2011hg}, or equivalently, the residual gauge symmetry of Chern-Simons gravity, which leads to Virasoro algebra. Note that from the anomaly one can determine the central charge:
\begin{equation}
    c=6\kc={3l\over 2G_N}\,.
\end{equation}
For the conformal transformation, one can integrate the infinitesimal transformation into a finite one
\begin{align}
	\mathcal{L}(z)=[f'(z)]^2\mathcal{L}_0- {\kc\over 4\pi } \left[ {f'''(z)\over f'(z)} -{3\over 2} \left({f''(z)\over f'(z) }\right)^2 \right]\,,
\end{align}
where $f(z)$ is the finite residual gauge transformation (conformal transformation) parameter. Also, noting that we have
\begin{equation}
	\mathcal{L}_0=-{ \pi \kc  \over  2\tau^2}\,,\label{eq: modular parameter and L0}
\end{equation}
one can write $\mathcal{L}(z)$ as
\begin{align}
	\mathcal{L}=&[f'(z)]^2\mathcal{L}_0- {\kc\over 4\pi } \left[ {f'''(z)\over f'(z)} -{3\over 2} \left({f''(z)\over f'(z) }\right)^2 \right]\cr
	=&-{\kc \over 4\pi} \text{Sch}\left[\tan \left({\pi f(z) \over \tau }\right) ; z \right]\,.
\end{align}
With a similar analysis for $\bar{a}_{\bar{z}}$, the on-shell action in~\eqref{eq: onshell action} is found to be
\begin{equation}
	I_{\ons}= -  {i \kc  \over 4\pi } \int {d^2 z \over \im (\tau) }\left[\tau \left( {2\pi^2 \over \tau^2 } [f'(z)]^2+\mbox{Sch}[f(z),z] \right) -\bar{\tau} \left( {2\pi^2 \over \bar{\tau}^2 }[\bar{f}'(\bar{z})]^2 +   \mbox{Sch}[\bar{f}(\bar{z}),\bar{z}]\right)  \right]\,,\label{eq: schwarzian action}
\end{equation}
which is analogous to the finite temperature Schwarzian action derived in~\cite{Maldacena:2016hyu}.

In the metric-like formulation of 3-dimensional gravity, one implements Dirichlet boundary conditions with the Gibbons-Hawking term, which leads to interaction between the holomorphic and anti-holomorphic soft modes. This interaction, however, is suppressed at large-$N$ (or at large central charges). On the other hand, in the Chern-Simons formalism, we did not include the Gibbons-Hawking term, and we have the decoupled on-shell action.

\subsection{Four point function in Euclidean correlators}
\label{sec: euclidean four pt function}

To evaluate the out-of-time-ordered correlator via analytic continuation, we start with the Euclidean four point function on the boundary. For simplicity, we consider cases where the four point function can be viewed as two point function of bi-local operators $\Phi(z_1,\bar{z}_1;z_2,\bar{z}_2)$ of which the leading contribution is one-point function (classical solution) of each bi-local operator
\begin{equation}
	\langle \Phi_1(z_1,\bar{z}_1;z_2,\bar{z}_2)\Phi_2(z_3,\bar{z}_3;z_4,\bar{z}_4)  \rangle=G_1(z_{12},\bar{z}_{12})G_2(z_{34},\bar{z}_{34})+\cdots \,,
\end{equation}
where $G_i(z_{12},\bar{z}_{12})$ ($i=1,2$) is defined by
\begin{equation}
    G_i(z_{12},\bar{z}_{12})\equiv \langle \Phi_i(z_1,\bar{z}_1;z_2,\bar{z}_2)\rangle\hspace{8mm} (i=1,2) \,.
\end{equation}
We now list some cases where such a bi-local operator can be considered.

\vspace{3mm}

\noindent
{\bf Heavy and light scalar operators with sparse spectrum:} One can consider the boundary-to-boundary four point function of heavy and light matter scalar fields. Then, the leading contribution will be factorized into a product of boundary-to-boundary two point functions of light and heavy operators, respectively. And, one may take the bi-local operators as follows
    \begin{equation}
        \Phi_H\sim \mathcal{O}_H\mathcal{O}_H\,,\qquad \Phi_L\sim \mathcal{O}_L\mathcal{O}_L\,.
    \end{equation}
 
\vspace{3mm}

\noindent   
{\bf Wilson line and Master field in the $sl(2)$ Vasiliev equation:} The Wilson line will play a role in pure Chern-Simons gravity. In particular, there has been extensive number of works on various Wilson lines in the context of higher spin gravity~\cite{Ammon:2013hba,deBoer:2013vca,Castro:2014tta,deBoer:2014sna,Castro:2018srf,Narayan:2019}. Here, we can consider the simplest (boundary-to-boundary) Wilson line as our bi-local operator
    \begin{equation}
        \Phi_{\text{\tiny Wilson}}(z_1,\bar{z}_1;z_2,\bar{z}_2)=\lim_{r\rightarrow \infty} e^{4h r}\tr\left[\mathcal{P}\exp\left(-\int^{r,z_1}_{r,z_2} A \right)\mathcal{P}\exp\left(-\int_{r,\bar{z}_1}^{r,\bar{z}_2} \bar{A} \right)\right]\,,
    \end{equation}
    where we take the $r$ coordinate of each end symmetrically. For constant connections $a$ and $\bar{a}$, this becomes a boundary-to-boundary propagator
    \begin{equation}
        \Phi_{\text{\tiny Wilson}}(z_1,\bar{z}_1;z_2,\bar{z}_2)\sim {1\over \left[\sin{2\pi z_{12}\over \tau }\sin {2\pi \bar{z}_{12}\over \bar{\tau} } \right]^{2h}   }\,.
    \end{equation}
For $sl(N)$ Chern-Simons gravity, the conformal dimension $h$ is negative
    \begin{equation}
        h=\bar{h}=-{N-1\over 2}<0\,.
    \end{equation}
    In fact, a similar object was observed in the Vasiliev equation~\cite{Hijano:2013fja}. In the $SL(N)$ version of the Vasiliev system, the equation of motion for the matter master field $C\in SL(N)$ is given by
    \begin{equation}
        dC + AC-C\bar{A}=0\,.
    \end{equation}
    For the given background connections $A$ and $\bar{A}$, we have
    \begin{align}
        C(r_1,z_1,\bar{z}_1;z_2,\bar{z}_2)=&\lim_{r_2 \rightarrow \infty} e^{2hr_2}\mathcal{P}\exp\left(-\int^{r_1,z_1}_{r_2,z_2} A \right)\tilde{c}_0\mathcal{P}\exp\left(-\int_{r_1,\bar{z}_1}^{r_2,\bar{z}_2} \bar{A} \right)\\
        =& b^{-1}(r_1)\mathcal{P}\exp\left(-\int^{z_1}_{z_2} a \right)c_0\mathcal{P}\exp\left(-\int_{\bar{z}_1}^{\bar{z}_2} \bar{a} \right)b^{-1}(r_1) \,,
    \end{align}
    where $\tilde{c}_0$ is a constant matrix. After taking $r_2\rightarrow \infty$ limit, it becomes ``the highest weight state'' \ie 
    \begin{equation}
        \lim_{r_2\rightarrow \infty} b(r_2)\tilde{c}_0 b^{-1}(r_2)=c_0 \,, \hspace{10mm}\mbox{where}\qquad (c_0)_{ij}\sim \delta_{i1}\delta_{j1} \,.
    \end{equation}
    It was shown~\cite{Ammon:2011ua,Hijano:2013fja} that the trace of the master field $C$ corresponds to the physical scalar field in the higher spin gravity. In particular, for the constant connection $a$ and $\bar{a}$, it becomes a bulk-to-boundary propagator~\cite{Hijano:2013fja}. Note that the coordinates $(z_2,\bar{z}_2)$ was chosen as the initial condition of the equation of motion, and it corresponds to the position of operator inserted on the boundary. The conformal dimension $h$ of the master field is the same as that of Wilson line
\begin{equation}
    h=\bar{h}=-{N-1\over 2}<0\label{eq: conformal dimension of scalar field}\,.
\end{equation}
This is not surprising in the higher spin AdS/CFT. In the $hs[\lambda]$ higher spin gravity which contains the infinite tower of higher spin field together with scalar field, the conformal dimension of the scalar field is given by 
\begin{equation}
    h={1\over 2}(1+\lambda) \,,
\end{equation}
where $\lambda\in [0,1]$. In the semi-classical limit where we take analytic continuation $\lambda\rightarrow -N$, one can truncate the infinite tower of higher spin field, and the gauge sector can be described by $SL(N)$ Chern-Simons gravity. In this limit, the conformal dimension of the scalar field becomes \eqref{eq: conformal dimension of scalar field}. Like Wilson line, one can also take the boundary-to-boundary propagator from the master field $\mathcal{C}$ as our bi-local operator:
\begin{equation}
    \Phi_{\text{\tiny Master}}(z_1,\bar{z}_1;z_2,\bar{z}_2)=\lim_{r_1\rightarrow \infty} e^{2h r_1} \tr [C(r_1,z_1,\bar{z}_1;z_2,\bar{z}_2)] \,.
\end{equation}

In the BTZ background the modular parameter $\tau,\bar{\tau}$ becomes purely imaginary, and the periodicity of $\phi$ requires the one point function of bi-local operator (which will be the boundary-to-boundary propagator) to be
\begin{equation}
	G(z_1,z_2;\bar{z}_1,\bar{z}_2)\equiv \sum_{m\in \mathbb{Z} } {1\over \left[\sin{\pi (z_{12} + 2\pi m )\over \tau }\right]^{2h} \left[\sin {\pi (\bar{z}_{12} + 2\pi m )\over \bar{\tau} } \right]^{2\bar{h}}  }\,.
\end{equation}
Note that for the case of Wilson line this corresponds to the summation of Wilson lines winding the non-contractible cycle of solid torus. For convenience, we define
\begin{equation}
	G_m(z_1,z_2;\bar{z}_1,\bar{z}_2)\equiv  {1\over \left[\sin{\pi (z_{12} + 2\pi m )\over \tau }\right]^{2h} \left[\sin {\pi (\bar{z}_{12} + 2\pi m )\over \bar{\tau} } \right]^{2\bar{h}}  }\,,
\end{equation}
and the boundary-to-boundary propagator can be written as
\begin{equation}
	G_{\text{\tiny BTZ}}=\sum_{m\in \mathbb{Z}}  G_m(z_1,z_2;\bar{z}_1,\bar{z}_2)\,.
\end{equation}

For the non-constant background $a$ and $\bar{a}$, our bi-local operator can be understood as a gravitationally dressed operator, and we can expand our bi-local operator  $\Phi(z_1,\bar{z}_1;z_2,\bar{z}_2)$ around that in the constant background $G(z_1,z_2;\bar{z}_1,\bar{z}_2)$
\begin{equation}
    \Phi^{\text{\tiny dressed}}(z_1,\bar{z}_1;z_2,\bar{z}_2)=G(z_1,\bar{z}_1;z_2,\bar{z}_2)+ \epsilon G^{(1)}(z_1,\bar{z}_1;z_2,\bar{z}_2)+\cdots \,,
\end{equation}
where $\epsilon$ is an infinitesimal expansion parameter. Recall that our non-constant solution is generated from the constant BTZ solution by an infinitesimal residual gauge transformation, which corresponds to a conformal transformation on the boundary. Hence, the expansion of the dressed operator can be understood as an infinitesimal conformal transformation of the bi-local operator on the boundary. Namely, under an infinitesimal conformal transformation
\begin{equation}
    z\qquad\Longrightarrow \qquad  f(z)=z+ \epsilon_n e^{-{2\pi i n z\over \tau}}\,,
\end{equation}
the bi-local operator can be written as
\begin{align}
    \Phi^{\text{\tiny dressed}}(z_1,\bar{z}_1;z_2,\bar{z}_2)=&[\partial f(z_1)]^{h}G(f(z_1),\bar{z}_1;f(z_2),\bar{z}_2)[\partial f(z_2)]^{h}\cr
    =& G(z_1,\bar{z}_1;z_2,\bar{z}_2)+ \epsilon_n \delta_{\epsilon_n} G^{\text{\tiny dressed}}(z_1,z_2;\bar{z}_1,\bar{z}_2) +\cdots \,,
\end{align}
where $\delta_{\epsilon_n} G_m(z_1,z_2;\bar{z}_1,\bar{z}_2)$ is found to be
\begin{align}
	{\delta_{\epsilon_n} G^{\text{\tiny dressed}}(z_1,z_2;\bar{z}_1,\bar{z}_2)\over G(z_1,z_2;\bar{z}_1,\bar{z}_2) }=- {4 i h \pi\over \tau}  e^{-{2\pi  i n \chi \over \tau} } \left[n\cos { 2\pi n \sigma\over \tau } -{\sin { 2\pi n\sigma\over \tau}\over \tan {2\pi (\sigma + \pi m )\over \tau }}\right]\,.	
\end{align}
Here, we used the center of bi-local coordinates and relative coordinates defined by
\begin{equation}
	\chi \equiv {1\over 2}(z_1+z_2)\,,\qquad \sigma\equiv {1\over 2}(z_1-z_2)\,,
\end{equation}
and a similar relation for the anti-holomorphic coordinates. As in~\cite{deMelloKoch:2018ivk}, one can also understand it as expansion of bi-local field to construct a bi-local conformal partial wave function. For this, one can consider correlation function of the bi-local field and $\epsilon_{-n}$, which is conjugate to $\epsilon_n$
\begin{equation}
    \langle \epsilon_{-n} \Phi^{\text{\tiny dressed}}(z_1,\bar{z}_1;z_2,\bar{z}_2)\rangle \sim \delta_{\epsilon_n} G(z_1,\bar{z}_1;z_2,\bar{z}_2)+\cdots \,.
\end{equation}
This is nothing but the conformal Ward identity. 

It is more clear to see the soft mode expansion of the gravitational dressed Wilson line (as well as the master field). The bi-local operator from the Wilson line with non-constant connection $a=h^{-1} a_\const h +h^{-1} \partial h$ can be written as follows
\begin{align}
        &\Phi^{\text{\tiny dressed}}_{\text{\tiny Wilson}}(z_1,\bar{z}_1;z_2,\bar{z}_2)\cr
        =&\lim_{r\rightarrow \infty} e^{-4|h| r}\tr\left[b^{-1}h^{-1}(z_1)e^{-(z_1-z_2)a_\const}h(z_2)[b(r)]^2h(z_2) e^{-(\bar{z}_2-\bar{z}_1)\bar{a}_\const}h^{-1}(z_1) b^{-1} \right] \,.
\end{align}
For infinitesimal residual gauge parameter $h(z)$, we expand $h(z)$ around identity
\begin{equation}
    h(z)=\mathbb{I}+ \lambda(z)+\cdots \,.
\end{equation}
Recall that the infinitesimal residual gauge transformation is parametrized by $\lambda_1(z)$. Then, from the mode expansion of $\lambda_1(z)$
\begin{equation}
    \lambda_1(z)=z+ \sum_{n\in \mathbb{Z}} \epsilon_n e^{-{2\pi i n z\over \tau}} \,,
\end{equation}
we can also expand the dressed Wilson line with respect to $\epsilon_n$'s
\begin{equation}
    \Phi^{\text{\tiny dressed}}_{\text{\tiny Wilson}}(z_1,\bar{z}_1;z_2,\bar{z}_2)=\Phi_{\text{\tiny Wilson}}(z_1,\bar{z}_1;z_2,\bar{z}_2)+ \sum_{n\in \mathbb{Z} }\epsilon_n [\delta_{\epsilon_n} \Phi^{\text{\tiny dressed}}_{\text{\tiny Wilson}}(z_1,\bar{z}_1;z_2,\bar{z}_2) ]+ \cdots \,.
\end{equation}

For the BTZ background, we also consider the expansion of the image of the dressed bi-local operator $\Phi^{\text{\tiny dressed}}_{m}(z_1,\bar{z}_1;z_2,\bar{z}_2)$ due to the periodicity of $\phi\sim \phi+2\pi$, and we denote it by $\tilde{f}_{n,m}$
\begin{align}
	\tilde{f}_{n,m}(\chi, \sigma;\bar{\chi},\bar{\sigma})\equiv &\delta_{\epsilon_n} \Phi^{\text{\tiny dressed}}_{m} \cr
	=&- {4 i h \pi\over \tau} G_m(z_1,z_2;\bar{z}_1,\bar{z}_2) e^{-{2\pi  i n \chi \over \tau} } \left[n\cos { 2\pi n \sigma\over \tau } -{\sin { 2\pi n\sigma\over \tau}\over \tan {2\pi (\sigma + \pi m )\over \tau }}\right]\,.	
\end{align}
In evaluating OTOCs, a particular configuration of $(z_1,z_2,z_3,z_4)$ simplify the four point function~\cite{Maldacena:2015waa,Maldacena:2016hyu,Yoon:2017nig,Narayan:2017hvh,Narayan:2019}. Namely, we will consider the configuration $(z_1,\bar{z}_1;z_2,\bar{z}_2)=(\chi-{\tau\over 4},\bar{\chi}-{\bar{\tau}\over 4};\chi+{\tau\over 4},\bar{\chi}+{\bar{\tau}\over 4} )$ or $(\chi,\sigma,\bar{\chi},\bar{\sigma})=(\chi,-{\tau\over 4},\bar{\chi},-{\bar{\tau}\over 4})$. For example, $G_m$ becomes
\begin{equation}
	G_m\left(\chi , -{\tau \over 4}; \bar{\chi},-{\bar{\tau} \over 4}\right)={1\over \left[\cos {2 \pi^2 m\over \tau}\right]^{2h}\left[\cos {2 \pi^2 m\over \bar{\tau} }\right]^{2\bar{h}} }\,,
\end{equation}
and $\tilde{f}_{n,m}$ can also be simplified as follows
\begin{equation}
	\tilde{f}_{n,m}\left(\chi , -{\tau \over 4}; \bar{\chi} ,-{\bar{\tau} \over 4}\right)=- {4 i h \pi\over \tau} {e^{- {2\pi i  n \chi\over \tau} } \over \left[\cos {2 \pi^2 m\over \tau}\cos {2 \pi^2 m\over \bar{\tau} }\right]^{2h} } \left[n\cos {  n \pi \over 2 } -\sin {  n \pi \over 2} \tan { 2\pi^2 m \over \tau }\right]\,.	
\end{equation}
Note that the second term is odd in $m$, which will be cancelled when we sum them up over the integers $m$. Hence, the expansion of the dressed bi-local field becomes
\begin{align}
	\delta_{\epsilon_n}\sum_m \Phi^{\text{\tiny dressed}}_{m}\left(\chi , -{\tau \over 4}; \bar{\chi},-{\bar{\tau} \over 4}\right)=& - {4 i h \pi\over \tau} e^{-{2\pi i  n \chi \over \tau} } n\cos {  n \pi \over 2 } \sum_{m\in \mathbb{Z}} {1 \over \left[\cos {2 \pi^2 m\over \tau}\right]^{2h}\left[\cos {2 \pi^2 m\over \bar{\tau} }\right]^{2\bar{h}} }\cr
	=&- {4 i h \pi\over \tau} e^{-{2\pi  i n \chi \over \tau} } n\cos {  n \pi \over 2 }G_{\text{\tiny BTZ}}\left(\chi, -{\tau \over 4}; \chi,-{\bar{\tau} \over 4}\right) \,. \label{eq: deviation of dressed operator}
\end{align}
Note that a special choice of the configuration allow us to factor out $G_{\text{\tiny BTZ}}(\chi, -{\tau \over 4}; \chi,-{\bar{\tau} \over 4})$. In general, this is not true, in particular, the same choice of coordinates does not simplify $\delta_{\epsilon_n} \Phi^{\text{\tiny dressed}}_{m}$ for the case of higher spin gravity~\cite{Narayan:2019}. Nevertheless, each term in the summation over $m$ has the same exponential growth in time. Hence, the summation would, at most, change the overall factor of the exponential growth, and therefore, could change the scrambling time.

Now, using the expansion of the dressed bi-local operator, we will evaluate the two point function of bi-local operators, which corresponds to the four point function of the local field. For this, we need to evaluate the correlation function of soft graviton modes, $\epsilon_n$'s. This can be evaluated from the on-shell action \eqref{eq: schwarzian action} of the soft graviton mode around the constant background. Since the on-shell action is non-linear, we evaluate the quadratic action by taking an infinitesimal fluctuation around the identity
\begin{align}
	f(z)=z+ \epsilon_n e^{-{ 2\pi i n z \over \tau } }\hspace{4mm},\hspace{8mm}\bar{f}(\bar{z} )=\bar{z}+ \bar{\epsilon}_n e^{-{ 2\pi i n \bar{z} \over \tau } } \,.
\end{align}
Expanding the on-shell action, the quadratic action takes the form
\begin{equation}
	I_{\ons}^{(2)}=  i \kc  \sum_{n\geqq 2 }  \left[  {16\pi^4\over \tau^3 } n^2(n^2-1) \epsilon_n\epsilon_{-n} -   {16\pi^4\over \bar{\tau}^3 } \bar{n}^2(\bar{n}^2-1) \bar{\epsilon}_n \bar{\epsilon}_{-n} \right] \,.
\end{equation}
Note that the on-shell action vanishes for $n=0,\pm 1$, which corresponds to the isometry of constant solutions.\footnote{For the BTZ black hole, the $SL(2,\mathbb{C})$ isometry of AdS vacuum is broken to $U(1)$,  due to the periodicity of $\phi$. However, since we consider the covering space of $\varphi$, we still have the $SL(2,\mathbb{C})$ isometry. On the other hand, the soft mode eigenfunction in the BTZ background reflects this symmetry breaking, \ie $\tilde{f}_{\pm1 ,m}\ne 0$ for $m\ne 0$.} From the quadratic on-shell action, one can read off the two point function of the boundary graviton fluctuation
\begin{equation}
	\langle \epsilon_n \epsilon_{-n} \rangle= {\kappa \over n^2 (n^2-1)}\,,\hspace{8mm}\langle \bar{\epsilon}_n \bar{\epsilon}_{-n} \rangle= -{\bar{\kappa} \over n^2 (n^2-1)}\,,
\end{equation}
where $\kappa$ and $\bar{\kappa}$ are defined by
\begin{equation}
	\kappa=  {\tau^3\over 16 \pi^4  i \kc }\,,\hspace{8mm} 	\bar{\kappa}\equiv  - {\bar{\tau}^5\over 16 \pi^4  i \kc }\,.
\end{equation}

We will evaluate the following configuration of the OTOC regularized by the thermal density matrix $\rho\equiv e^{-\beta H}$:
\begin{equation}
    F(t,\varphi)= \mathfrak{tr} \left[\rho^{1\over 4} V(0)\rho^{1\over 4} W(t,\varphi)\rho^{1\over 4} V(0)\rho^{1\over 4} W(t,\varphi) \right] \,.
\end{equation}
For this, we consider the Euclidean four point function at
\begin{align}
	(z_1,\bar{z}_1)=&\left(\chi-{\tau\over 4},\bar{\chi}- {\bar{\tau}\over 4} \right)\,, \\
	(z_2,\bar{z}_2)=&\left(\chi+{\tau\over 4},\bar{\chi}+{\bar{\tau}\over 4} \right)\,, \\
	(z_3,\bar{z}_3)=&(0, 0 ) \,, \\
	(z_4,\bar{z}_4)=&\left({\tau\over 2}, {\bar{\tau}\over 2} \right) \,.
\end{align}
Recalling the expansion of the dressed bi-local operator in~\eqref{eq: deviation of dressed operator} at the special configuration, one can further simplify that of ($z_3,\bar{z}_3;z_4,\bar{z}_4$)
\begin{equation}
	{\delta G_{\text{\tiny BTZ}}({\tau \over 4} , -{\tau \over 4}; {\bar{\tau} \over 4},-{\bar{\tau} \over 4})\over  G_{\text{\tiny BTZ}}({\tau \over 4} , -{\tau \over 4}; {\bar{\tau} \over 4},-{\bar{\tau} \over 4}) }=- {4 i h \pi\over \tau} (-1)^{n\over 2} n\cos {  n \pi \over 2 } \,.
\end{equation}
Finally, the contribution of the boundary graviton fluctuation to the Euclidean four point function can be evaluated as follows\footnote{Here, and in the following, we use the number 1 to denote $(z_1,\bar{z}_1)$, the number 2 to denote $(z_2,\bar{z}_2)$, and so on.}
\begin{align}
	&{F(1,2,3,4 )\over G_{\text{\tiny BTZ}}(1,2)G_{\text{\tiny BTZ}}(3,4)}\cr
	=&1+\sum_{|n|\geqq 2}   (-1)^{n\over 2} n^2 \cos^2{  n \pi \over 2 } \left[{16h^2 \pi^2 \over \tau^2 } e^{-{2\pi  i n \chi \over \tau} } \langle \epsilon_n \epsilon_{-n} \rangle+{16\bar{h}^2 \pi^2 \over \bar{\tau}^2 } e^{-{2\pi  i n \bar{\chi} \over \bar{\tau} } } \langle \bar{\epsilon}_n \bar{\epsilon}_{-n} \rangle\right]\cr
	=&1+\sum_{|n|= 2,4, 6,\cdots}  \left[{16h^2 \pi^2 \over \tau^2 }\kappa {(-1)^{n\over 2} e^{-{2\pi  i n \chi \over \tau} }\over n^2-1 }+ {16\bar{h}^2 \pi^2 \over \bar{\tau}^2 }\bar{\kappa} {(-1)^{n\over 2} e^{-{2\pi  i n \bar{\chi} \over \bar{\tau} } }\over n^2-1 }  \right]\,,
\end{align}
where only the even terms give a contribution to the four point function due to the $\cos {n\pi \over 2}$ factor. Then, one can rewrite the summation as a contour integral where the contour $\mathcal{C}$ is a collection of circles centered at $n=2 \mathbb{Z}$ with small radius. 
\begin{align}
	{F(1,2,3,4 )\over G_{\text{\tiny BTZ}}(1,2)G_{\text{\tiny BTZ}}(3,4)}=1&+{16h^2 \pi^2 \over \tau^2 }\kappa  {1\over 2\pi i} \oint_{\mathcal{C} } d\zeta {{\pi\over 2} \over \sin {\pi \zeta\over 2} }{ e^{-{2\pi  i \zeta \chi \over \tau}  }\over \zeta^2-1 }\cr
	&+{16\bar{h}^2 \pi^2 \over \bar{\tau}^2 }\bar{\kappa}  {1\over 2\pi i} \oint_{\mathcal{C} } d\zeta {{\pi\over 2} \over \sin {\pi \zeta\over 2} }{ e^{-{2\pi  i \zeta \bar{\chi} \over \bar{\tau} }  }\over \zeta^2-1 }\,.
\end{align}
By pushing the contour to infinity, we pick up the simple pole at  $\zeta=-1,0,1$, and the contour integral along $\mathcal{C}$ becomes the residue at $\zeta=-1,0,1$
\begin{align}
	{F(1,2,3,4 )\over G_{\text{\tiny BTZ}}(1,2)G_{\text{\tiny BTZ}}(3,4)}=1 & -{16h^2 \pi^2 \over \tau^2 }\kappa   \left[{\pi \over 4} e^{-{2\pi  i \chi \over \tau}  } - 1 +{\pi  \over 4}  e^{{2\pi  i  \chi \over \tau}} \right]\cr
	&-{16\bar{h}^2 \pi^2 \over \bar{\tau} ^2 }\kappa   \left[{\pi \over 4} e^{-{2\pi  i \bar{\chi} \over \bar{\tau}}  } - 1 +{\pi  \over 4}  e^{{2\pi  i  \bar{\chi} \over \bar{\tau} }} \right]\,.\label{eq: euclidean four point function}
\end{align}
%

\subsection{Boyer-Lindquist co-rotating frame}
\label{sec: BL}

In this section, we will shortly discuss the {\it Boyer-Lindquist co-rotating frame}, which is necessary for a consistent analytic continuation from Euclidean time to Lorentzian time in the four point functions.

In~\cite{Maldacena:2001kr}, the TFD state is interpreted as Hartle-Hawking state generated by an Euclidean time translation of size $\beta/2$. Recall that the TFD state in~\eqref{def: tfd state} can be viewed as a state generated by an $\beta/2$ Euclidean time evolution with the new twisted hamiltonians $\mathcal{H}_{L}$ and $\mathcal{H}_R$
\begin{equation}
	\widetilde{H}_{L/R}\equiv H_{L/R} +\Omega \mathcal{J}_{L/R} \,,
\end{equation}
where $H_{L/R}$ and $\mathcal{J}_{L/R}$ is the hamiltonian and angular momentum operator, respectively. The form of TFD state in~\eqref{def: tfd state} implies that the Hartle-Hawking state will be constructed by an Euclidean time evolution of size $\beta/2$ with the Killing vector $\widetilde{H}=\partial_t +{1\over \Omega}\partial_\phi$.

Note that the Killing vector $\widetilde{H}$ becomes null at horizon (the tip in the Euclidean BTZ black holes), which leads to a smooth tip of the Euclidean rotating BTZ black hole (and smooth bifurcation surface in Lorentzian geometry). Had we chosen other Killing vector to construct the Hartle-Hawking state, for example, $\widetilde{H}$ , we would not have a smooth bifurcation surface. In this sense, the twisted Hamiltonian $\widetilde{H}$ is more natural to construct the Hartle-Hawking state of the rotating BTZ black hole (assuming that this Hartle-Hawking state exists).

After constructing the Hartle-Hawking state, one can evolve the state in Lorentzian time either by $H$ or by $\widetilde{H}$, and they lead to different interpretation in the TFD formulation.

If (to obtain Lorentzian time correlators) we perform the analytic continuation of the Euclidean correlators, which are evolved by twisted Hamiltonian, the resulting correlator will agree with the case where the whole complex time contour (\eg TFD time contour or Schwinger-Keldysh time contour) is evolved by $\widetilde{H}_L$ or $\widetilde{H}_R$. Hence, we define a new time coordinate $\tilde{t}$ generated by $\widetilde{H}_L$ and a new angular coordinate $\tilde{\phi}$ orthogonal to the Killing vector $\mathcal{H}_L$
\begin{align}
	\tilde{t}_E\equiv {1\over 2} \left(t_E + {1\over \Omega } i \varphi\right) \,,\hspace{6mm}
	\tilde{\phi}\equiv {1\over 2} (\varphi +\Omega i t_E)\,,
\end{align}
or
\begin{align}
	\tilde{t}\equiv {1\over 2} \left(t + {1\over \Omega }  \varphi\right)  \,,\hspace{6mm}
	\tilde{\varphi}\equiv {1\over 2} (\varphi - \Omega  t)\,.
\end{align}
%
%
%
%
In our TFD formulation, the operators are defined on the space $(\tilde{t},\tilde{\phi})$, where $\tilde{t}$ denotes the complex time coordinates. Then, the periodicity condition becomes
\begin{equation}
	(\tilde{t}_E,\tilde{\phi})\sim (\tilde{t}_E - i \beta, \tilde{\phi}) \,.
\end{equation}
%

\subsection{OTOCs from analytic continuation}
\label{sec: analytic continuation to otoc}

From the Euclidean four point function~\eqref{eq: euclidean four point function}, we consider the following term which will grow exponentially after the analytic continuation to real time
\begin{align}
	{F(1,2,3,4 ) \over G_{\text{\tiny BTZ}}(1,2)G_{\text{\tiny BTZ}}(3,4)} = 1 &- {  l h^2 \over 2(r_+ +r_-)\kc  } \exp \left(-{2\pi i \over \tau} z\right)\cr
	&- {  l h^2 \over  2(r_+ - r_-)\kc  } \exp \left(-{2\pi i \over \bar{\tau} } \bar{z}\right)+\cdots \,.  \label{eq: euclidean four point function final}
\end{align}
Here, we used
\begin{align}
	-{4h^2 \pi^3 \over \tau^2 }\kappa = 
	- {  l h^2 \over  2(r_+ +r_-)\kc  }\hspace{3mm},\hspace{8mm}
	-{4\bar{h}^2 \pi^3 \over \bar{\tau}^2 }\bar{\kappa} 
	=- {  l \bar{h}^2 \over  2(r_+  - r_-)\kc  } \,.
\end{align}
As we discussed, it is natural to use the ``Boyer-Lindquist co-rotating'' frame for the analytic continuation to the real time because the Lorentzian time is generated by the twisted Hamiltonian. In term of Boyer-Lindquist co-rotating coordinates, the holomorphic and anti-holomorphic coordinates $z, \bar{z}$ are written as follows
\begin{align}
	z=&\varphi+ i{t_E\over l}
	={r_++r_-\over r_-}\tilde{\phi} +i {r_+ - r_-\over r_+} {t_E\over l }\,, \cr 
	\bar{z}=&\varphi - i{t_E\over l} 
	=- {r_+ - r_-\over r_-}\tilde{\phi} - i {r_+ + r_-\over r_+} {t_E\over l }  \,.
\end{align}
Therefore, we perform the analytic continuation of the Euclidean four point function in~\eqref{eq: euclidean four point function final} within the Boyer-Lindquist co-rotating frame
\begin{align} \label{main44}
	{F(1,2,3,4 ) \over G_{\text{\tiny BTZ}}(1,2)G_{\text{\tiny BTZ}}(3,4)}  
	=&1 - {  l h^2 \over  \pi^2(r_+ +r_-)\kc  } \exp \left[-{(r_++ r_-)^2 \over lr_-}\tilde{\phi}-{ r_+^2 -r_-^2 \over l^2 r_+}i\tilde{t}_E \right] \cr
	&\hspace{5mm}- {  l h^2 \over  \pi^2(r_+ - r_-)\kc  }\exp \left[-{(r_+- r_-)^2 \over lr_-}\tilde{\phi}-{ r_+^2 -r_-^2 \over l^2 r_+}i\tilde{t}_E \right]\cr
	\Longrightarrow \quad& 1 - {  l h^2 \over  \pi^2(r_+ + r_-)\kc  }  \exp \left[-{(r_++ r_-)^2 \over lr_-}\tilde{\phi}+{ r_+^2 -r_-^2 \over l^2 r_+}\tilde{t} \right] \cr
	&\hspace{5mm} - {  l h^2 \over  \pi^2(r_+ - r_-)\kc  } \exp \left[-{(r_+- r_-)^2 \over lr_-}\tilde{\phi}+{ r_+^2 -r_-^2 \over l^2 r_+}\tilde{t} \right] \,.
\end{align}
Hence, one can read off the Lyapunov exponent and the butterfly velocities in the Boyer-Lindquist co-rotating frame
\begin{align}
	\lambda_L=&{2\pi \over \beta}\,,\\
	v_B^+=& { r_-\over l r_+ } {r_+- r_- \over r_+ + r_- }= \Omega {1-l\Omega \over 1+l\Omega}\,,\\
	v_B^-=& { r_-\over l r_+ } {r_+ +  r_- \over r_+ - r_- }=\Omega {1+l\Omega \over 1- l\Omega}\,.
\end{align}
In terms of the Schwarzschild coordinates, the long time behavior of the OTOC can be expressed as
\begin{align}
    {F(1,2,3,4 ) \over G_{\text{\tiny BTZ}}(1,2)G_{\text{\tiny BTZ}}(3,4)}= &1 - {  l h^2 \over  \pi^2(r_+ + r_-)\kc  }  \exp \left[{r_++r_-\over l^2}(t-l\varphi) \right] \cr
	&\hspace{5mm} - {  l h^2 \over  \pi^2(r_+ - r_-)\kc  } \exp \left[{r_+-r_-\over l^2}(t+l\varphi) \right] \,. \label{main33}
\end{align}
Hence, in  Schwarzschild coordinates, both modes have the same butterfly velocity, $v_B/c=1$, but different Lyapunov exponents
\be
\lambda_{L}^{\pm}=\frac{r_{+} \pm r_{-}}{\ell}=\frac{2\pi}{\beta}\frac{1}{1\mp \ell \Omega}\,.
\ee
This result perfectly matches the result obtained with the gravity eikonal approximation in section \ref{sec: eikonal approximation}.

\section{Discussion}
\label{sec: discussion}

In this paper we have studied the onset of chaos in rotating BTZ black holes. We have computed OTOCs using two different methods: the elastic eikonal gravity approximation \cite{Shenker:2014cwa}, and a new method that is based on the Chern-Simons formulation of 3-dimensional gravity. 
In the first case the OTOC can be obtained from a high energy shock wave collision in the bulk, while in the second method one explicitly derives an effective Schwarzian-like action for the boundary degrees of freedom, and computes the OTOC from analytic continuation of the Euclidean 4-point function. Both methods give the same result, which, in terms of Schwarzschild coordinates, takes the form
\be \label{main22}
\text{OTOC}(t,\varphi_{12}) \approx 1+ C_1 e^{\frac{2\pi}{\beta_{+}}\left(t+\ell \varphi_{12}\right)}+C_2 e^{\frac{2\pi}{\beta_{-}}\left(t-\ell \varphi_{12} \right)}\,,\,\,\,\,\beta_{\pm}=\beta (1\mp \ell \Omega)\,,
\ee
with $ \ell \,\Omega =\frac{r_-}{r_+}$ being the chemical potential for angular momentum. 
The two terms correspond to left and right moving modes, which have the same butterfly velocity\footnote{Here $c=1/\ell$ is the boundary speed of light.}, $v_B/c=1$, but different Lyapunov exponents
\be
\lambda_L^{\pm}=\frac{2\pi}{\beta_{\pm}}=\frac{2\pi}{\beta}\frac{1}{1\mp \ell \Omega}\,.
\ee
This result is consistent with \cite{Poojary:2018esz}, where the author used the metric formulation of 3-dimensional Euclidean gravity to derive an effective action for the boundary degrees of freedom of the rotating BTZ black hole. A somewhat similar result was also found in \cite{Stikonas:2018ane}, where the author studied the disruption of the mutual information in TFD states. However, due the special configuration considered in \cite{Stikonas:2018ane}, their result is only sensitive to the mode with the lowest Lyapunov exponent ($\lambda_L=\frac{2\pi}{\beta_-}$).

Naively, since
$\lambda_L^{-} \leq \frac{2\pi}{\beta} \leq \lambda_L^{+}$, the above result seems to indicate that one of the Lyapunov exponents is less than maximal, while the one violates the chaos bound. However, in the decomposition described in section \ref{sec: tfd of rotating btz black hole}, the parameters $\beta_{\pm}^{-1}$ are precisely the effective temperatures of the left and right moving modes. Therefore, the chaos bound is different for each mode, and our result shows that each mode saturates its own chaos bound. This interpretation is very natural if one views the BTZ black hole as a part of a D1-D5 brane system. From this perspective, $\beta(1\mp \ell \Omega)$ are the inverse temperatures of the left and right moving excitations on the effective string \cite{Maldacena:1998bw}.

\subsubsection*{Co-rotating frame}
The black hole temperature, $\beta^{-1}$, is defined in co-rotating coordinates $(t,\phi)$, in terms of which the result takes the form
\be
\text{OTOC}(t,\phi_{12}) \approx 1+ C_1 e^{\frac{2\pi}{\beta}\left(t+\frac{\ell \phi_{12}}{1-\ell \, \Omega} \right)}+C_2 e^{\frac{2\pi}{\beta}\left(t-\frac{\ell \phi_{12}}{1+\ell \, \Omega} \right)}\,,
\label{eq-OTOC-eikonal}
\ee
In this frame, both modes have the same temperature, which results in the same Lyapunov exponent, namely $\lambda_L=\frac{2\pi}{\beta}$. This is consistent with \cite{Reynolds:2016pmi}, in which the authors (using the co-rotating frame) studied chaos in the rotating BTZ black hole using the geodesic approximation. Since the authors of \cite{Reynolds:2016pmi} only consider homogeneous shock waves, their result has no angular dependence, and it is controlled by a single mode, with maximal Lyapunov exponent.

The change of frame breaks the symmetry between the butterfly velocities, which now become
\be
\frac{v_B^{\pm}}{c}=1\pm \ell \Omega\,.
\ee
 The rotation affects the butterfly velocities, and one of them becomes superluminal when $\Omega > 0$. Since $v_B$ defines an effective light cone for the butterfly effect \cite{Roberts:2014isa}, one expects this quantity to be bounded by the speed of light. In fact, in asymptotically AdS geometries satisfying the Null Energy Condition (NEC) one can show that \cite{Qi:2017ttv}
\be
v_B \leq c\,.
\label{eq-boundVB}
\ee
This bound is known to be violated when the boundary theory displays non-local effects \cite{Fischler:2018kwt}, which are related to a non-asymptotically AdS geometry in the dual gravitational description. Since the rotating BTZ geometry is asymptotically AdS and satisfies NEC, the violation of the bound (\ref{eq-boundVB}) {in this setup must have a different reason. We think this violation occurs because the rotation breaks the $\mathrm{Z}_2$ isometry of the non-rotating geometry, and this introduces an asymmetry between the left and right moving modes in the co-rotating frame. This is reminiscent of the cases studied in \cite{Jahnke:2017iwi,Giataganas:2017koz,Avila:2018sqf}, where the breaking of rotational symmetry caused the violation of a different (and more strong) bound for the butterfly velocity.}

{
Finally, we note that, at $r_{-}=0$, we should recover the Rindler AdS$_3$ result of S\&S \cite{Shenker:2014cwa} by taking the limit $\beta/\ell <<1$ and $\phi <<1$ of our results\footnote{We thank Mark Mezei for calling our attention to this.}. Indeed, using that\footnote{See (4.6) of \cite{Fu:2018oaq}.}
\bea
C_1^{-1}& \propto &\frac{2r_{+}}{\ell}\left(e^{2\pi (r_+ + r_-)/\ell} -1\right)\,, \nonumber\\
C_2^{-1}& \propto &\frac{2r_{+}}{\ell}\left(1-e^{-2\pi (r_+-r_-)/\ell} \right)\,,
\eea 
one can easily check that, at $r_-=0$, the OTOC becomes
\be
\text{OTOC}(t,\phi) \propto \frac{\ell}{2r_{+}} \frac{1}{1-e^{-2\pi r_+/\ell}}\left( e^{-\frac{r_{+}}{\ell}\phi}+e^{-\frac{2\pi r_+}{\ell}} e^{\frac{ r_{+}}{\ell}\phi} \right) \,,
\ee
where the first term matches the S\&S result, while the second term ($\sim e^{\frac{ r_{+}}{\ell}\phi}$) is exponentially suppressed\footnote{Note that $\beta/\ell=2\pi \ell/r_+ <<1$ implies $r_+/\ell >> 1$. So, the second term is small because of the multiplicative factor $e^{-\frac{2\pi r_+}{\ell}}$. The first term, $e^{-\frac{r_+}{\ell}\phi}$, is not small because $\phi <<1$.} in the limit $\beta/\ell <<1$ and $\phi <<1$. }

\subsubsection*{Boyer-Lindquist co-rotating frame}

In the Boyer-Lindquist coordinates, the OTOC takes the form
\be
\text{OTOC}(\tilde{t},\tilde{\varphi}_{12})=1+ B_1\, \exp\left[ \frac{2\pi}{\beta}\left(\tilde{t}-\frac{ \tilde{\varphi}_{12}}{v_B^-}\right)\right]+B_2\, \exp\left[ \frac{2\pi}{\beta}\left(\tilde{t}+ \frac{ \tilde{\varphi}_{12}}{v_B^+}\right)\right]\,,
\label{eq-OTOC-BL}
\ee
where the butterfly velocities are given by
\be
	v_B^+ = \Omega {1-l\Omega \over 1+l\Omega}\,,\,\,\,\,v_B^-=\Omega {1+l\Omega \over 1- l\Omega}\,.
\ee
In this frame both modes have the same, maximal, Lyapunov exponent, but they have different butterfly velocities.
Just like in the case of co-rotating coordinates, one of the butterfly velocities can become superluminal, i.e., $v_B > c$. But here this only happens when $\Omega > \sqrt{2}-1$, while in the co-rotating coordinates this happens for any value of $\Omega$.

\subsubsection*{Extremal limit}
The extremal limit occurs when $r_{-} \rightarrow r_{+}$. In the co-rotating frame (or in the Boyer-Lindquist co-rotating frame), both modes have zero Lyapunov exponent. That is consistent with the idea that the co-rotating observer only sees the zero-temperature black hole.
Interestingly, this is not the case in the non-rotating frame. In this case, $\lambda_L^{-}=0$, but $\lambda_{L}^{+}=r_{+}/\ell^2$. That means that one of the modes survives in the extremal limit, with effective temperature given by $T=\frac{r_{+}}{2\pi \ell^2}$. A similar phenomenon was also observed in \cite{Caputa:2013lfa}.

\acknowledgments
It is a pleasure to thank Ioannis Papadimitriou, Miok Park, Nakwoo Kim, and Juan Pedraza for extensive discussions, and Rohan R. Poojary and Mark Mezei for useful correspondence. We also thank A. Misobuchi for helpful discussions about the rotating BTZ black hole, and Yongjun Ahn for carefully reviewing the section on the Eikonal approximation. The work of VJ and KK was supported in part by Basic Science Research Program through the National Research Foundation of Korea~(NRF) funded by the Ministry of Science, ICT \& Future Planning~(NRF2017R1A2B4004810) and GIST Research Institute(GRI) grant funded by the GIST in 2019. JY thanks the Erwin Schrodinger International Institute~(ESI) where this work was completed during the program ``Higher Spins and Holography 2019'', and JY would like to thank the organizers for giving an opportunity to present our work.


\appendix

\section{Finite residual gauge transformations}
\label{app: finite transf}

In this appendix, we will consider a finite residual gauge transformation for the constant connection $a_\const$. An analogous calculation holds for $\bar{a}_\const$. Let us start with
\begin{equation}
	a_z= L_1 - {2\pi \over \kc} \mathcal{L}_0 L_{-1} = \begin{pmatrix}
	0 & {2\pi \over \kc} \mathcal{L}_0 \\
	1 & 0 \\
	\end{pmatrix} \,.
\end{equation}
Under the finite gauge transformation by $h$, we have
\begin{equation}
	a=h^{-1}(z) (d + a_\const ) h(z) \,, \hspace{7mm}\mbox{where}\quad  h \in SL(2) \,.
\end{equation}
By the Gauss decomposition of $h(z)$ 
\begin{equation}
	h(z)= e^{h_1(z) L_1} e^{h_0(z) L_0} e^{h_{-1}(z) L_{-1}} = \begin{pmatrix}
	1 & 0 \\
	h_1(z) & 0\\
	\end{pmatrix}\begin{pmatrix}
	e^{h_0(z)\over 2} & 0 \\
	0 & e^{-{h_0(z)\over 2}}\\
	\end{pmatrix}\begin{pmatrix}
	1 & -h_{-1}(z) \\
	0 & 0\\
	\end{pmatrix}\ ,
\end{equation}
one can express $h_0$ and $h_{-1}$ in terms of $h_1$ by demanding that the gauge transformation by $h(z)$ keeps the gauge condition in~\eqref{eq: gauge condition}:
\begin{equation}
	a= L_1 - {2\pi \over \kc}\mathcal{L} L_{-1} = \begin{pmatrix}
	0 & {2\pi \over \kc}\mathcal{L}(z)\\
	1 & 0\\
	\end{pmatrix}\,.
\end{equation}
We obtain
\begin{align}
	e^{-h_0(z)}=&  1-{2\pi \over \kc} \mathcal{L}_0 [h_1(z)]^2 +h'_1(z)  \,, \\
	h_{-1}(z)=&\ -{2\pi \over \kc} \mathcal{L }_0 h_1(z)  - {1 \over 2} h_0(z)'  \,,
\end{align}
and $\mathcal{L}[h_1]$ is a complicated functional of $h_1$. It is better to reparametrize $h_1(z)$
by defining $\phi(z)$ via  
\begin{equation}
	h_1(z)= - {1\over \sqrt{-{2\pi \over \kc}\mathcal{L}_0} } \tan \left[\sqrt{-{2\pi \over \kc}\mathcal{L}_0}\left( t -  \phi(z)\right) \right]\,.
\end{equation}
In this case $\mathcal{L}(z)$ becomes the finite temperature Schwarzian
\begin{equation}
	\mathcal{L}(z)= -{\kc\over 4\pi } \left[ {\phi'''\over \phi'} -{3\over 2} \left( {\phi''\over \phi'} \right) \right]+  \mathcal{L}_0 (\phi')^2\,.
\end{equation}
From \eqref{eq: modular parameter and L0}, the gauge parameter $h(z)$ can be obtained to be 
\begin{equation}
    h(z) = \begin{pmatrix}
	\cos({\pi \over \tau} (z- \phi(z))) &  {\pi \over \tau} \sin({\pi \over \tau} (z- \phi(z)))\\
	-{ \sin({\pi \over \tau} (z- \phi(z))) \over {\pi \over \tau}} & \cos({\pi \over \tau} (z- \phi(z)))\\
	\end{pmatrix} 
	\begin{pmatrix}
	{1 \over \phi'(z)^{1 \over 2} } &  - {\phi''(z) \over 2 \phi'(z)^{3 \over 2} } \\
	0 & \phi'(z)^{1 \over 2}\\
	\end{pmatrix}\,.
\end{equation}



\bibliographystyle{JHEP}

\providecommand{\href}[2]{#2}\begingroup\raggedright\endgroup

\end{document}